\pdfoutput=1  
\documentclass[fleqn,usenatbib,useAMS]{mnras}

\usepackage{newtxtext,newtxmath}

\usepackage[T1]{fontenc}
\usepackage{ae,aecompl}

\usepackage{graphicx}	
\usepackage{amsmath}	
\usepackage{amssymb}	
\usepackage[noabbrev]{cleveref}     
\usepackage{booktabs}               
\usepackage{bm}                     
\usepackage{subcaption}             
\newcommand{\PolyChord}{\texttt{PolyChord}}     
\newcommand{\MultiNest}{\texttt{MultiNest}}     
\newcommand{\dyPolyChord}{\href{https://github.com/ejhigson/dyPolyChord}{\textcolor{black}{\texttt{dyPolyChord}}}}

\newcommand{\nestcheck}{\href{https://github.com/ejhigson/nestcheck}{\textcolor{black}{\texttt{nestcheck}}}}
\newcommand{\fgivenx}{\texttt{fgivenx}}
\newcommand{\numrepeats}{\texttt{num\_repeats}}
\newcommand{\Z}{\ensuremath{\mathcal{Z}}}
\renewcommand{\d}[1]{\ensuremath{\operatorname{d}\!{#1}}}
\newcommand{\e}{\mathrm{e}}

\title[Bayesian sparse reconstruction]{Bayesian sparse reconstruction: a brute-force approach to astronomical imaging and machine learning}

\author[E. Higson et al.]{%
Edward Higson,$^{1,2}$\thanks{E-mail: e.higson@mrao.cam.ac.uk}
Will Handley,$^{1,2}$
Michael Hobson$^{1}$
and Anthony Lasenby$^{1,2}$
\\
$^{1}$Astrophysics Group, Battcock Centre, Cavendish Laboratory, JJ Thomson Avenue, Cambridge CB3 0HE, UK\\
$^{2}$Kavli Institute for Cosmology, Madingley Road, Cambridge, CB3 0HA, UK
}

\date{Accepted XXX\@. Received YYY\@; in original form ZZZ}

\pubyear{2015}

\volume{{\rm in press}}
\begin{document}\label{firstpage}
\pagerange{\pageref{firstpage}--\pageref{lastpage}}
\maketitle

\begin{abstract}
We present a principled Bayesian framework for signal reconstruction, in which the signal is modelled by basis functions whose number (and form, if required) is determined by the data themselves.
This approach is based on a Bayesian interpretation of conventional sparse reconstruction and regularisation techniques, in which sparsity is imposed through priors via Bayesian model selection.
We demonstrate our method for noisy 1- and 2-dimensional signals, including astronomical images.
Furthermore, by using a product-space approach, the number and type of basis functions can be treated as integer parameters and their posterior distributions sampled directly.
We show that order-of-magnitude increases in computational efficiency are possible from this technique compared to calculating the Bayesian evidences separately, and that further computational gains are possible using it in combination with dynamic nested sampling.
Our approach can also be readily applied to neural networks, where it allows the network architecture to be determined by the data in a principled Bayesian manner by treating the number of nodes and hidden layers as parameters.
\end{abstract}
\begin{keywords}
methods: statistical ---
methods: data analysis ---
methods: numerical ---
techniques: image processing
\end{keywords}



\section{Introduction}

Sparse signal processing and Bayesian inference are both well-established methods for data analysis, and have a considerable amount in common.
However, these two approaches are often considered somewhat distinct from one another, and this is often reflected in the relatively small overlap of the communities who develop and apply each technique.
Nevertheless Bayesian interpretations of sparse signal processing techniques have been pursued by a number of authors in the signal processing community --- see for example \citet{Ji2008}, \citet{Tipping2001} and \citet{Wipf2004}.
In addition, Bayesian inference with imposed sparsity has been applied to astronomical problems \citep[such as in][]{Warren2017,Sciacchitano2018a,Jones2018}.

In this paper we outline a principled Bayesian approach for simultaneously imposing sparsity and performing dictionary learning to determine the optimal basis set for representing the signal, and discuss how Bayesian inference provides a very natural framework for sparsity.
In our method a signal is modelled as the superposition of a set of basis functions, whose number and form are determined by the data themselves.
Sparsity can be imposed directly via the prior on the number of basis functions $N$, while simultaneous dictionary learning is performed through the estimation of parameters describing the location and shape of the basis functions.

The optimum number of basis functions $N$ with which to model a signal can be determined using Bayesian model selection by calculating the Bayesian evidence for each value.
However it is equivalent (and often more computationally efficient and convenient) to treat $N$ as an integer parameter, and sample directly from the joint posterior of $N$ and the other parameters describing the $N$ basis functions.
The final inference may then be obtained by either by choosing the maximum {\em a posteriori\/} value of $N$ or, better, by marginalising over $N$ to give a multi-model solution \citep{Parkinson2013} with the fit for each number of basis functions weighted by its posterior probability.
This method can be further generalised to select from a variety of types of basis functions $T$ (such as Gaussians, Fourier modes, wavelet families, shapelets, etc.), with the full version involving inference over the joint space $T$, $N$ and the basis functions' parameters.

While our principled approach is computationally expensive, we show that it is practical in low data regime using current numerical methods at reasonable computational cost (see \Cref{tab:core_hours} in Appendix~\ref{app:core_hours} for details of the number of core hours used to produce our results).
In addition, this paper is intended as a proof of principle for applications where our method is not currently feasible but will be made so in the future by advances in numerical methods and increases in computational power.

The paper proceeds as follows: \Cref{sec:bsr_background} describes standard regression techniques, regularisation and sparsity.
\Cref{sec:bayesian_approach} then provides a Bayesian perspective on these topics --- including introducing our formulation of ``Bayesian sparse reconstruction'' and a discussion of how it can be implemented numerically.
\Cref{sec:1d_bsr,sec:2d_bsr} demonstrate applying our approach to 1- and 2-dimensional signal processing, including of astronomical images from the Hubble Space Telescope eXtreme Deep Field \citep{Illingworth2013}.
In addition, in \Cref{sec:nn}, we apply the Bayesian sparse reconstruction framework to artificial neural networks, where we perform Bayesian inference over the joint space of network architectures and network parameters by treating the number of nodes and (if required) the number of hidden layers as parameters.

\section{Regression, regularisation and sparsity}\label{sec:bsr_background}

We begin by presenting some background on standard approaches to regression, regularisation and sparsity.
This provides context for the Bayesian framework presented in \Cref{sec:bayesian_approach}, in which all these methods may be reinterpreted.
This section is intended to draw out the common themes in numerous popular signal reconstruction methods, and describe them in a unified manner.

\subsection{Standard non-parametric regression}\label{sec:regression}

Regression involves using data points $\{\bm{x}_d,y_d\}$ (including random noise) to reconstruct some function $y = f(\bm{x};\bm{\theta})$, where $\bm{\theta}$ is some number of free parameters and the semicolon separates variables from parameters.
Such inverse problems are common in science, and are typically ill-posed\footnote{An ``ill-posed'' problem does not satisfy all three of the conditions for a problem to be ``well-posed'' outlined by \citet{Hadamard1902}. The conditions are: a solution exists, the solution is unique and the behaviour of the solution changes continuously with changes in the parameters and data.}.
For example, in monochrome image reconstruction each data point is a pixel with 2-dimensional (centre) position $\bm{x}$ and scalar intensity value $y$.
In general $\bm{x}$ and $y$ can be vectors of any dimension; for simplicity in this paper we consider only scalar outputs $y$, but the results easily generalise to vector outputs.

\begin{figure}
	\centering
    \includegraphics[width=\linewidth]{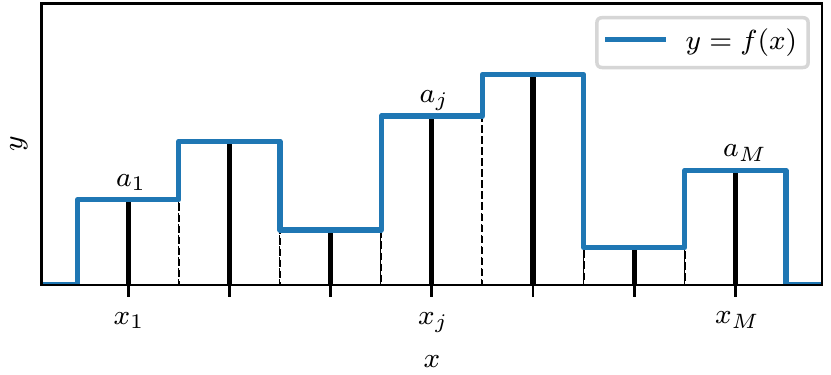}
    \caption{Free-form decomposition of a 1-dimensional function $y=f(x)$ into $M$ pixels with amplitudes (free parameters) $\bm{\theta}=(a_1, a_2, \dots, a_M)$.}\label{fig:freeform_bar}
\end{figure}

When a good model for the data is not available {\em a priori}, a traditional non-parametric approach is to use a {\em free-form\/} solution \citep{Sivia2006} in which the function is pixelated and the value at each pixel is fitted.
This is a standard way of performing ``brute-force'' numerical calculations on computers, and is equivalent to fitting a delta function (or more accurately ``top-hat'') basis function centred on each of the $M$ pixels with their amplitudes as free parameters, giving $M$ degrees of freedom.
Ironically, such ``non-parametric'' approaches thus contain many parameters --- typically far more than ``parametric'' approaches.
The free form approach is illustrated for 1-dimensional input $x$ in \Cref{fig:freeform_bar} \citep[which is based on Figure 6.1 of][]{Sivia2006}, but can be performed in arbitrary dimensions.

\begin{figure}
	\centering
    \includegraphics[width=\linewidth]{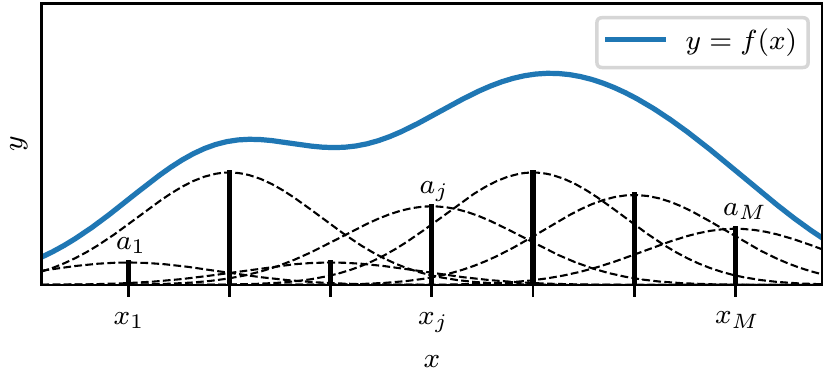}
    \caption{Free-form decomposition of a 1-dimensional function $y=f(x)$ into $M$ Gaussian basis functions with standard deviation $\sigma$, each centred on a pixel, with amplitudes (free parameters) $\bm{\theta}=(a_1, a_2, \dots, a_M)$.}\label{fig:freeform_gau}
\end{figure}

Smoothness can be encoded into the solution by replacing the delta functions with broader basis functions $\phi(\bm{x};\bm{x}_j,\sigma)$, with fixed centres $\bm{x}_j$ located on each of the $M$ pixels and their width determined by a shared shape parameter $\sigma$.
For Gaussian basis functions:
\begin{equation}
    f(\bm{x};\bm{a}, \sigma) = \sum_{j=1}^{M} a_j \phi(\bm{x}; \bm{x}_j, \sigma) =  \sum_{j=1}^{M} a_j \exp \left( -\frac{{|\bm{x}-\bm{x}_j|}^2}{2\sigma^2} \right).\label{equ:fixed_basis_funcs}
\end{equation}
This is illustrated for a 1-dimensional input $x$ in \Cref{fig:freeform_gau} \citep[based on Figure 6.7 of][]{Sivia2006}.

\subsection{Optimisation and regularisation}\label{sec:regularisation}

\begin{figure}
	\centering
    \includegraphics{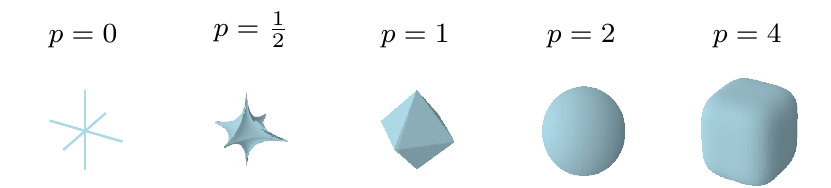}
    \caption{$L_p=1$ surfaces in 3 dimensions for different values of $p$, with $L_0$ representing the number of non-zero components of the vector.}\label{fig:pnorm}
\end{figure}

The values of the parameters $\bm{\theta}$ are typically chosen by minimising the squared $L_2$ norm of the differences between the model and the data (also referred to as the squared residuals or $\chi^2$).
Here the $L_p$ norm of a vector is defined for $p > 0$ as ${\lVert \bm{v} \rVert}_p \equiv {(\sum_i {\lvert v_i \rvert}^p )}^{1/p}$ and the $L_0$ norm is the number of non-zero components; this is illustrated for different values of $p$ in \Cref{fig:pnorm}.
The squared $L_2$ norm approach yields the maximum likelihood estimate (MLE) for $\bm{\theta}$ under certain restrictive conditions,\footnote{These include that the residuals on each data point must be independently normally distributed, and that there are no errors in the independent variables $\bm{x}$.} although it is commonly applied when these are not met; see \citet[][Chapter 8]{Sivia2006} for a more detailed discussion and recommended modifications to the least squares procedure for different types of data.

When using the squared $L_2$ norm, the optimisation is
\begin{equation}
    \min_{\bm{\theta}}
    \sum_{d=1}^D {\left( y_d - f(\bm{x}_d; \bm{\theta}) \right)}^2
    =
    \min_{\bm{\theta}}
    {\left\lVert \bm{y} - \bm{\hat{y}} \right\rVert}_2^2,\label{equ:chisquared}
\end{equation}
where $\bm{y} = \{ y_1, \dots, y_D \}$ are the data values and $\bm{\hat{y}} = \{ f(\bm{x}_1), \dots, f(\bm{x}_D) \}$ are the fit values.
For simplicity we assume for the moment that the shape of the basis functions is fixed and only the amplitudes are free parameters, in which case
\begin{equation}
    \min_{\bm{\theta}}
    {\left\lVert \bm{y} - \bm{\hat{y}} \right\rVert}_2^2
    =
    \min_{\bm{a} \in \mathcal{R}^M}
    {\left\lVert \bm{y} - \Phi \bm{a} \right\rVert}_2^2,\label{equ:chisquared_a}
\end{equation}
where the vector $\bm{a} = (a_1,a_2,\dots,a_N)$ determines the basis functions' amplitudes and $\Phi=(\phi_1,\phi_2,\dots,\phi_N)$ is a $D\times M$ basis matrix.

Typically a regularisation term is added to penalise more complex models; this is to prevent the analysis fitting noise in the data set and producing a result which will not generalise to new data sets (``overfitting'').
Some popular choices are:
\begin{itemize}
    \item The (squared) $L_2$ norm --- used in the Wiener filter \citep{Wiener1949} and ridge regression \citep{Hoerl1970}:
        \begin{equation}
            \qquad \min_{\bm{a} \in \mathcal{R}^M} {\left\lVert \bm{y} - \Phi \bm{a} \right\rVert}_2^2 + \lambda {\rVert \bm{a} \lVert}_2^2.\label{equ:l2_reg}
        \end{equation}
    \item The $L_1$ norm --- used in the Lasso \citep{Tibshirani1996}, compressed sensing and for imposing sparsity:
        \begin{equation}
            \qquad \min_{\bm{a} \in \mathcal{R}^M} {\left\lVert \bm{y} - \Phi \bm{a} \right\rVert}_2^2 + \lambda {\rVert \bm{a} \lVert}_1.\label{equ:l1_reg}
        \end{equation}
    \item The $L_0$ norm --- used in matching pursuit \citep{Mallat1993}, iterative thresholding \citep{Elad2007}, compressed sensing and for imposing sparsity:
        \begin{equation}
            \qquad \min_{\bm{a} \in \mathcal{R}^M} {\left\lVert \bm{y} - \Phi \bm{a} \right\rVert}_2^2 + \lambda {\rVert \bm{a} \lVert}_0,\label{equ:l0_reg}
        \end{equation}
        where ${\rVert \bm{a} \lVert}_0$ simply counts the number of non-zero elements in the amplitude vector $\bm{a}$.
    \item The entropy --- used in the maximum entropy method (MEM):
        \begin{align}
            \qquad &\min_{\bm{a} \in \mathcal{R}^M} {\left\lVert \bm{y} - \Phi \bm{a} \right\rVert}_2^2 - \lambda S(\bm{a}),\label{equ:mem_reg}\\
            &S(\bm{a})=\sum_{i=1}^N a_i - m_i - a_i \ln\left(\frac{a_i}{m_i}\right),
        \end{align}
        where $m_i$ is a (model) amplitude value assigned to each basis function \citep{Ables1974,Gull1978}.
\end{itemize}

In principle, such optimisations define a ``solution curve'' $\hat{\bm{a}}(\lambda)$.
To obtain a particular solution one must choose a value for the regularisation parameter $\lambda$, which determines the relative importance of the accuracy of the fit to the data and the value of the regularising function; it is often chosen {\em a priori\/} but can be determined using heuristics or cross-validation.
For example, the regularisation constant for MEM has historically been chosen so the residual statistic equals its expectation value --- i.e.\ so $\chi^2 = D$ where $D$ is the number of data points \citep{Sivia2006}.
A more modern approach is to choose the value of $\lambda$ which maximises the Bayesian evidence (see \Cref{sec:Bayesian_inference}); this can also be used to select quantities such as the width $\sigma$ of the basis functions shown in \Cref{fig:freeform_gau} \citep{Sivia2006}.

It is worth noting that \cref{equ:l2_reg,equ:l1_reg,equ:l0_reg,equ:mem_reg} refer to the {\em synthesis\/} formulation which optimises over the parameters $\bm{a}$.
An alternative is the {\em analysis\/} approach in which the optimisation is performed directly with respect to the (vectorised) function $f(\bm{x})$, and $\bm{a}$ is replaced in \cref{equ:l2_reg,equ:l1_reg,equ:l0_reg,equ:mem_reg} with $\Phi^{-1}f(\bm{x})$.
This technique is commonly used in radio interferometry --- see for example \citet{Maisinger2004}, \citet{McEwen2011} and \citet{Cai2017a}.

\subsection{Sparse representations}\label{sec:compressed_sensing}

\begin{figure}
	\centering
    \includegraphics[width=\linewidth]{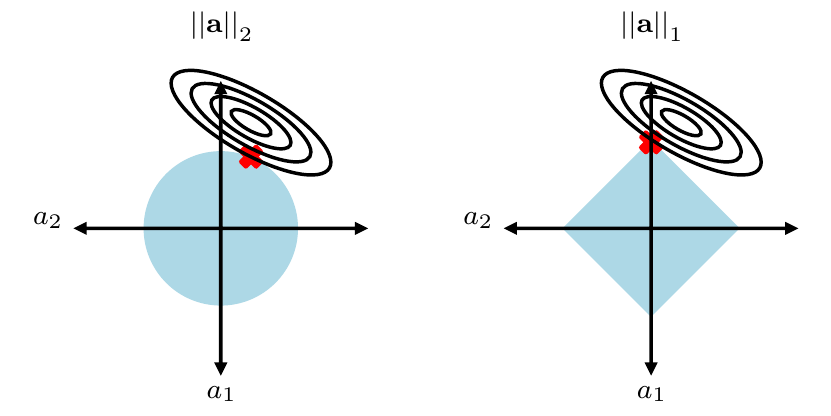}
    \caption{Illustration of $L_p$ norms promoting sparsity when $p<2$.
    The blue circle on the left plot shows the region of the parameter space ($a_1,a_2$) with an $L_2$ norm less than some maximum value, and the blue diamond on the right plot shows the region with an $L_1$ norm less than some maximum value.
    The black contours on each plot show an objective function to be optimised.
    Due to its angular shape, the region constrained by a maximum $L_1$ norm is more likely to have its maximum value of the objective function at a coordinate where one of the parameters is zero than the $L_2$ region.}\label{fig:pnorm_sparsity}
\end{figure}

In many practical signal and image processing applications we can use prior knowledge that the physical signals have ``sparse'' representations in which they have very few non-zero components (a low $L_0$ norm).
For example, astronomical images with many pixels can often be well represented by a relatively small number of point sources or wavelets.
Sparse solutions are promoted by choosing a regularisation term $L_p$ with $p<2$, in which case $L_p$ surfaces have singular points at sparse solutions \citep{Bach2012}; this is illustrated graphically in \Cref{fig:pnorm_sparsity}.

Sparsity is key to {\em compressed sensing\/}: a popular signal processing technique for efficiently recovering high-dimensional vector signals under the assumption that they are sparse in some basis \citep[see][for a detailed discussion]{Eldar2012}.
Sparse solutions can be found by $L_0$-optimisation, but this is computationally challenging and is non-convex, meaning standard convex optimisation cannot be used.
The success of compressed sensing is based on approximating the $L_0$-norm with the $L_1$-norm --- the smallest $p$ for which the $L_p$ norm is convex.
Compressed sensing has been applied successfully to a variety astronomical problems; see for example \citet{Bobin2008} and \citet{Wiaux2009}.

\subsection{Adaptive basis functions and dictionary learning}

In order to find representations for data sets which are sparse (use relatively few basis functions) we now generalise the reconstructions described in~\eqref{sec:regression} by allowing each basis function's location and shape to be determined by parameters $\bm{p}_i$ and fitted to the data.
The signal is reconstructed as 
\begin{equation}
    f(\bm{x};\bm{a},\bm{p}_1,\dots,\bm{p}_N) = \sum_{i=1}^{N} a_i \phi(\bm{x};\bm{p}_i),
    \label{equ:ad_basis_sum}
\end{equation}
where now the number of basis functions $N$ can easily be much smaller than the number of pixels $M$.

For a given data set, some types of basis function will provide more natural and sparse representation than others.
We can further generalise~\eqref{equ:ad_basis_sum} using parameterised dictionary learning, by fitting different families of standard basis functions (determined by a categorical variable $T$).
The optimisation then determines $T$, as well as each basis function's amplitude $a_i$ and parameters $\bm{p}_i$ by reconstructing the signal as
\begin{equation}
    f(\bm{x};T,\bm{a},\bm{p}_1,\dots,\bm{p}_N) = \sum_{i=1}^{N} a_i \phi^{(T)}(\bm{x};\bm{p}_i).
\end{equation}
Commonly used basis function families include Gaussians, wavelets and shapelets.

\section{A Bayesian approach}\label{sec:bayesian_approach}

We now consider the Bayesian interpretation of the regression and regularisation discussed in \Cref{sec:bsr_background}, before outlining our approach to Bayesian sparse reconstruction and how it can be performed computationally.

\subsection{Background: Bayesian inference}\label{sec:Bayesian_inference}

Bayesian inference (see \citet{Sivia2006} or \citet{MacKay2003} for a detailed discussion) can be divided into {\em parameter estimation\/} and {\em model comparison}.  
Given a model $\mathcal{M}$, inferences about its parameters $\bm{\theta}$ from data $\mathcal{D}$ can be made by calculating the posterior distribution of $\bm{\theta}$ using Bayes' theorem \citep{Bayes1763}:
\begin{equation}
    P(\bm{\theta}|\mathcal{M},\mathcal{D})
    =
    \frac{P(\mathcal{D}|\bm{\theta},\mathcal{M})P(\bm{\theta}|\mathcal{M})}{P(\mathcal{D}|\mathcal{M})}
    \equiv
    \frac{\mathcal{L}(\bm{\theta}) \pi(\bm{\theta})}{\Z},
    \label{equ:parameter_estimation}
\end{equation}
where $\mathcal{L}$, $\pi$, and $\mathcal{Z}$ are the likelihood, prior and Bayesian evidence respectively.
The evidence $\Z$ is a normalisation constant, and is computed by averaging the likelihood $\mathcal{L}$ over the prior $\pi$:
\begin{equation}
    \Z \equiv P(\mathcal{D}|\mathcal{M})
    =
    \int \mathcal{L}(\bm{\theta}) \pi(\bm{\theta})\d{\bm{\theta}}.
    \label{equ:Z_definition}
\end{equation}
Bayes' theorem can also be used to compare different models $\mathcal{M}_1,\mathcal{M}_2,\dots$ and assess which best describes the data.
The posterior probability of a given model is
\begin{equation}                                          
    P(\mathcal{M}_j|\mathcal{D})=\frac{P(\mathcal{D}|\mathcal{M}_j)P(\mathcal{M}_j)}{P(\mathcal{D})}=\frac{\Z_j \Pi_j}{\sum_k \Z_k \Pi_k},
    \label{equ:model_comparison}
\end{equation}
where $\Pi_j \equiv P(\mathcal{M}_j)$ denotes the prior probability of each model and the denominator of the final term sums over all competing models.
The evidence $\Z$ penalises more complex models so this approach naturally includes Occam's razor and favours sparse solutions.
Models may also be compared by computing log posterior odds ratios
\begin{equation}
	\mathcal{P}^j_k
    \equiv
    \log \left(\frac{P(\mathcal{M}_j|\mathcal{D})}{P(\mathcal{M}_k|\mathcal{D})}\right) = \log\left(\frac{\Z_j}{\Z_k}\right) + \log\left(\frac{\Pi_j}{\Pi_k}\right).
    \label{equ:Bayes_factor}
\end{equation}
where the ratio of evidences $\mathcal{B}^j_k = \Z_j / \Z_k$ is called a Bayes factor and is often used for comparison when the prior probability of the different models is not easily determined.
The Bayes factors do, however, depend on the priors on the models' parameters $\pi(\bm{\theta}_{\mathcal{M}_j})$ through the calculation of $\Z_j$ from~\eqref{equ:Z_definition}.
If the prior on different models is uniform, the Bayes factors are equal to the posterior odds ratios.

\subsection{Bayesian formulation of regression and regularisation}

Before introducing our full Bayesian sparse reconstruction framework in \Cref{sec:bsr}, we first give a Bayesian formulation of the regression and regularisation problems discussed in \Cref{sec:regression,sec:regularisation} as the comparison is very informative.
In these cases the number, type and shape of basis functions are fixed and the only parameters of the model are the amplitudes --- i.e. $\bm{\theta} = \bm{a}$.

In general, defining the likelihood of the basis function fit given some data $\mathcal{D}$ requires knowledge of how measurement errors are distributed.
For example, a common assumption in the literature is that there are independent Gaussian errors on the signal values $\{y_d \}$, and no errors on the data points' coordinates $\{\bm{x}_d\}$.
In this case the likelihood of the data given the model is
\begin{align}
\begin{split}
    \mathcal{L}(\bm{a})
    =
    P(\mathcal{D}|\bm{a},\mathcal{M})
    =&
    \prod_{d=1}^D
    \frac{1}{\sqrt{2\pi\sigma_y^2}}
    \exp\left(-\frac{{(y_d-f(\bm{x}_d;\bm{a}))}^2}{2\sigma_y^2}\right)
    \\
    \propto&
    \exp \left( - \frac{{\left\lVert \bm{y} - \Phi \bm{a} \right\rVert}_2^2}{2 \sigma^2_y} \right),
    \label{equ:fitting_likelihood}
\end{split}
\end{align}
recovering the (exponentiated) least squares objective function from~\eqref{equ:chisquared_a}.
Of course this assumption may be inappropriate for some data sets.
For example, for low-level photon-counting measurements a Poisson likelihood function (or similar) may be required.
Although the Bayesian formulation can naturally accommodate other likelihood functions, for simplicity we will henceforth consider only independent Gaussian errors.

In order for the Bayesian approach to give the same maximum {\em a posteriori\/} parameter values as the optimisation in~\Cref{sec:regularisation}, the prior $\pi(\bm{a})$ must correspond to the exponential of the regularisation terms in~(\ref{equ:l2_reg}-\ref{equ:mem_reg}).
For a more formal derivation of this result in the context of the Wiener filter, see \citet{Hobson1998} and \citet{Lasenby2001}.

In the Bayesian framework, the different regularisation techniques in~(\ref{equ:l2_reg}-\ref{equ:mem_reg}) are analogous to the following choices of priors:
\begin{itemize}
    \item $L_2$ (squared) regularisation~\eqref{equ:l2_reg} corresponds to a Gaussian prior:
        \begin{equation}
            \qquad \pi(\bm{a}) \propto \exp(- \lambda {\rVert \bm{a} \lVert}_2^2).\label{equ:l2_reg_pi}
        \end{equation}
    \item $L_1$ regularisation~\eqref{equ:l1_reg} corresponds to a Laplacian prior:
        \begin{equation}
            \qquad \pi(\bm{a}) \propto \exp(- \lambda {\rVert \bm{a} \lVert}_1).\label{equ:l1_reg_pi}
        \end{equation}
    \item $L_0$ regularisation~\eqref{equ:l0_reg} corresponds to an exponential prior on the number of non-zero components of $\bm{a}$:
        \begin{equation}
            \qquad  \pi(\bm{a}) \propto \exp(- \lambda {\rVert \bm{a} \lVert}_0) = \exp(-\lambda N), \label{equ:l0_reg_pi}
        \end{equation}
        where $N$ is the number of basis functions used in the signal reconstruction.
    \item Entropy regularisation~\eqref{equ:mem_reg} corresponds to an entropic prior
        \begin{equation}
            \qquad  \pi(\bm{a}) \propto \exp(\lambda S(\bm{a})),
            \label{equ:mem_reg_pi}
        \end{equation}
        where $S(\bm{a})$ is defined in~\eqref{equ:mem_reg}.
\end{itemize}
More generally other priors can be used.
For example, one may promote sparsity by using any prior which has fatter tails than a Gaussian and is also more concentrated at zero --- such priors prefer to shrink amplitudes to zero while also being lenient in allowing larger amplitudes.
Thus, as an alternative to the Laplacian distribution~\eqref{equ:l1_reg_pi}, one could use for example a Cauchy distribution
\begin{equation}
    \pi(\bm{a}) = \prod_{j=1}^M \frac{\lambda}{\pi} \frac{1}{\lambda^2 + a_j^2}.
\end{equation}

In addition, the $L_2$ regularisation prior~\eqref{equ:l2_reg_pi} can be generalised to include some covariance matrix $\bm{C}$, which may be a function of some further parameters $\bm{\theta}$,
\begin{equation}
    \pi(\bm{a}) \propto \exp(- \lambda \bm{a}^\intercal \bm{C}^{-1} \bm{a}).\label{equ:l2_reg_pi_gp}
\end{equation}
This form can be used to reconstruct a signal as a Gaussian process \citep[see][for an introduction]{Rasmussen2004}, with $\bm{C}$ representing its correlation structure.
When performing the optimisation, the basis matrix $\Phi$ most naturally contains Fourier modes.
The optimisation is typically performed by selecting both $\bm{\theta}$ and $\lambda$ to maximise the Bayesian evidence (sometimes the value of $\lambda$ is chosen {\em a priori\/}).
Indeed, Gaussian processes could be further generalised by using a different form for the prior term --- for example ``entropic processes'' with $\pi(\bm{a}) \propto \e^{-\lambda S (\bm{L}\bm{a})}$, where $S$ is defined as in~\eqref{equ:mem_reg} and $\bm{C}=\bm{L}\bm{L}^\intercal$ is the Cholesky decomposition of the signal correlation matrix \citep{Hobson1998}.

\subsection{Sampling and model selection}

Maximum {\em a posteriori\/} estimates of the parameters $\bm{a}$ can be found from the posterior distribution $\propto \mathcal{L}(\bm{a})\pi(\bm{a})$ in an analogous manner to the optimisations in~(\ref{equ:chisquared_a}-\ref{equ:mem_reg}).
However, a major advantage of the Bayesian approach is that it provides a {\em generative\/} model and allows the full posterior distribution to be sampled.
This provides additional information such as posterior distributions on the weights $\bm{a}$ and other quantities of interest.

Furthermore, the posterior distribution allows the appropriate number of basis functions to be chosen via Bayesian model selection by calculating posterior odds ratios~\eqref{equ:Bayes_factor}.
This naturally penalises more complex models and, with an appropriate choice of priors, provides a principled Bayesian method for creating models with the level of complexity which is justified by the data.
Finally one can either choose the maximum {\em a posteriori\/} number of basis functions or, better, marginalise over $N$ so the fit with each number of basis functions is weighted in proportion to its posterior probability.
Any {\em a priori\/} expectation of the degree of sparsity can be included in the priors, and there is no need for an additional regularisation term.

\subsection{Bayesian sparse reconstruction}\label{sec:bsr}

Following the discussion in the previous sections, we propose reconstructing the relationship $y=f(\bm{x})$ as a sum of $N$ basis functions $\phi^{(T)}$ of type $T$ with weights $a_i$ and shape and location parameters $\bm{p}_i$ as
\begin{equation}
    f(\bm{x};T, N, \bm{a}, \bm{p}_1,\dots,\bm{p}_N) = \sum_{i=1}^{N} a_i \phi^{(T)}(\bm{x},\bm{p}_i).\label{equ:basis_fit}
\end{equation}
One can then perform Bayesian inference over the full parameter space of $\bm{\theta} = (T, N, \bm{a}, \bm{p}_1,\dots,\bm{p}_N)$.

This approach has the desirable properties that:
\begin{itemize}
    \item full posterior distributions on parameters can be recovered by sampling (rather than simply optimising);
    \item the Bayesian calculations naturally penalise over-complex models. In addition, when there is an {\em a priori\/} justification, sparsity can be further enforced directly through priors on the total number of basis functions $N$;
    \item there are a variable number of basis functions with variable positions;
    \item there is no need to choose a regularisation constant $\lambda$;
    \item families and/or shapes of basis functions are determined and can be marginalised over;
    \item arbitrary constraints can be imposed on the reconstruction (not just positivity);
    \item any type of noise can be included --- e.g. Gaussian, Poisson, etc. If the size or nature of the noise is unknown, it can be expressed in terms of additional parameters which can be marginalised over;
    \item missing and/or irregular data can be accommodated;
    \item the model is generative and can easily be extended to deconvolution.
\end{itemize}

The remainder of this section discusses how Bayesian sparse reconstructions can be computed, with numerical tests presented in the following section.

\subsection{Vanilla and adaptive methods}\label{sec:adaptive}

Given some noisy signal to be reconstructed, Bayesian model selection can be used to determine an appropriate type $T$ and number $N$ of basis functions to use by calculating the Bayesian evidence $\Z_{T,N}$ for the fit using each combination (model) $T,N$.
Using~\eqref{equ:model_comparison}, the posterior probability of each model is proportional to $\Z_{T,N} \Pi_{T,N}$, where $\Pi_{T,N}$ is the prior probability of the model and over-complex models are penalised by lower evidences.
We term this the {\em vanilla method}.
One can then either select the model with the highest posterior probability or, better, use a combination of all models weighted by their posterior probability (``multi-model analysis'').

The {\em adaptive method\/} is an alternative product-space approach, which analyses a ``meta-model'' containing one or more discrete parameters with values corresponding to each individual model.
The likelihood of a sample is found by selecting the model indicated by the discrete parameters, then working out the likelihood for this model using the remaining parameters.
A fixed dimensionality which is sufficient for the individual model with the most parameters is used; for models with fewer parameters, the likelihood is independent of the remaining unneeded parameters.
This is an alternative to transdimensional sampling methods such as reversible-jump MCMC \citep{Green1995}.
\citet{Hee2016a,Hee2017} used the adaptive method in reconstructing 1-dimensional signals by linearly interpolating between $N$ points (``nodes''), with their co-ordinates as free parameters.
We generalise this approach by letting the integer parameter $N$ represent the number of basis functions (of any dimension) to be used, and when needed also including a second integer parameter $T$ to determine the form of the basis functions.
Posterior distributions of $T$ and $N$ are found using parameter estimation.

\subsection{Practical considerations for sampling the posterior}

The posterior is typically of moderate to large dimensionality, and will be non-convex and multimodal with pronounced degeneracies.
Furthermore, due to the integer parameters $T$ and $N$, methods requiring gradients cannot be used.
We explore the posterior using nested sampling \citep{Skilling2006}, which is well suited to such problems and can be performed using software packages such as \MultiNest{} \citep{Feroz2008,Feroz2009,Feroz2013} or \PolyChord{} \citep{Handley2015a,Handley2015b}.
The adaptive method calculates posterior odds ratios indirectly by sampling the integer parameters $T$ and $N$, and as a result its sampling errors have different properties to those of the vanilla method (which uses evidence calculations).
For a detailed discussion of sampling errors in nested sampling parameter estimation, see \citet{Higson2017a}.

Nested sampling calculations can be made significantly more computationally efficient (or alternatively more accurate for the same amount of computation) using dynamic nested sampling \citep{Higson2017b}.
In particular, dynamic nested sampling gives large efficiency gains for parameter estimation, meaning it works well with the adaptive method.
In contrast the efficiency gains for evidence calculations are relatively modest (except in low dimensions), so dynamic nested sampling only produces small speedups for model selection via the vanilla method and we do not use it in this case.
Results in this paper were calculated using \dyPolyChord{} \citep{Higson2018dypolychord} --- a dynamic nested sampling package based on \PolyChord.
Due to the challenging multimodal posteriors produced by the integer parameter in the adaptive method, we use a large fraction (50\%) of the total computational budget for each calculation on \dyPolyChord{}'s initial exploratory run.
This reduces the possible efficiency gain, but \dyPolyChord{} is still able to produce significant speedups compared to standard nested sampling.

\section{Fitting 1-dimensional data}\label{sec:1d_bsr}

We first demonstrate Bayesian sparse reconstruction by finding the dependence of some scalar quantity $y$ on another scalar variable $x$, and to make the example more challenging we allow errors on both the data values $y_d$ and positions $x_d$.

If each measurement has an independent error distribution $P(x_d,y_d|X_d,Y_d)$ about its true value $X_d,Y_d$ then the probability of the observed data given some set of true values is
\begin{equation}
    P(\mathcal{D}|\{X_d,Y_d\}) = \prod_{d=1}^D P(x_d,y_d|X_d,Y_d).
\end{equation}
The unknown true data values $X_d,Y_d$ are then marginalised out using the basis fitting model by taking $Y_d = f(X_d;T, N, \bm{a}, \bm{p}_1, \dots, \bm{p}_N)$ and integrating over the distribution of the $x$ coordinates at which data points were sampled $P(X_d)$.
Hence each likelihood call involves an integral for every data point and
\begin{multline}
    P(\mathcal{D}|P(X_d),T, N, \bm{a}, \bm{p}_1, \dots, \bm{p}_N) =
    \\
    \prod_{d=1}^D \int P(x_d,y_d|X_d,f(X_d)) \, P(X_d) \, \d{X_d},\label{equ:data_dist}
\end{multline}
where for brevity we have omitted the dependence of $f(X_d;T, N, \bm{a}, \bm{p}_1, \dots, \bm{p}_N)$ on the parameters $T,N,\bm{a}, \bm{p}_1, \dots, \bm{p}_N$.

We first consider samples to be taken uniformly in the range $X_- < X_d < X_+$ with independent Gaussian $x$ and $y$ errors of size $\sigma_{x}$ and $\sigma_{y}$. In this case~\eqref{equ:data_dist} gives the likelihood \citep{Hee2016a}
\begin{multline}
    \mathcal{L}(T, N, \bm{a}, \bm{p}_1, \dots, \bm{p}_N)
    =
    P(\mathcal{D}|T, N, \bm{a}, \bm{p}_1, \dots, \bm{p}_N)
    =\\                                                                 
    \prod_{d=1}^D \int_{X_-}^{X_+} \frac{\exp\left[-\frac{{(x_d-X_d)}^2}{2\sigma_x^2}-\frac{{(y_d-f(X_d))}^2}{2\sigma_y^2}\right]}{2\pi\sigma_x\sigma_y(X_+-X_-)} \d{X_d}.
    \label{equ:fitting_likelihood_hee}
\end{multline}

The priors on the parameters and models can be specified as required, and together with the likelihood can be used to sample numerically from the posterior and calculate evidences.
As $f(x;T,N, \bm{a}, \bm{p}_1, \dots, \bm{p}_N)$ is typically invariant under interchange of basis function index the prior space can be shrunk by a factor of $N$! by enforcing ordering using ``forced identifiability'' (sorted) priors; see \citet[][Appendix A2]{Handley2015b} for a more detailed discussion.

\subsection{Basis functions}

\begin{figure}
\begin{minipage}{.48\linewidth}
	\centering
    \includegraphics[width=\linewidth]{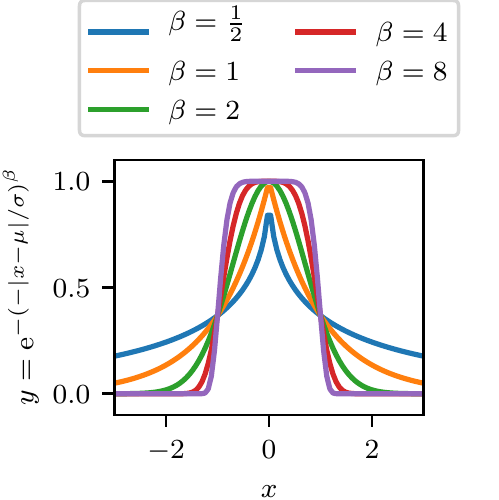}
    \subcaption{Generalised Gaussians~\eqref{equ:gg_1d} for different values of $\beta$; each has $\mu=0$ and $\sigma=1$.}%
    \label{fig:gg_demo}
\end{minipage}
\hfill
\begin{minipage}{.48\linewidth}
	\centering
    \includegraphics[width=\linewidth]{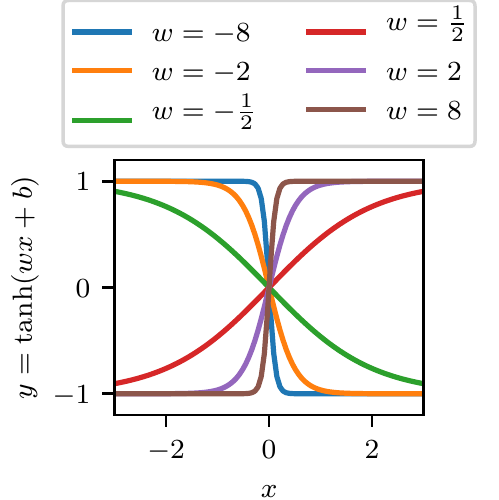}
    \subcaption{tanh functions~\eqref{equ:ta_1d} with different values of $w$. All the lines shown use $b = 0$.}%
    \label{fig:ta_demo}
\end{minipage}
\caption{Illustrations of 1-dimensional basis functions.}\label{fig:demo}
\end{figure}

The likelihood~\eqref{equ:fitting_likelihood_hee} applies for mixture models with any 1-dimensional basis function; we demonstrate it using 1-dimensional generalised Gaussians
\begin{equation}
    \phi^{(\mathrm{g1d})}(x;\bm{p}) = \phi^{(\mathrm{g1d})}(x,\mu,\sigma,\beta) = \e^{-{(|x - \mu|/\sigma)}^{\beta}}
    \label{equ:gg_1d}
\end{equation}
and 1-dimensional tanh functions
\begin{equation}
    \phi^{(\mathrm{t1d})}(x,\bm{p}) = \phi^{(\mathrm{t1d})}(x, w, b) = \tanh(w x + b).
    \label{equ:ta_1d}
\end{equation}
Their shape and location are determined by parameters $\bm{p} = (\mu,\sigma,\beta)$ and $\bm{p}=(w,b)$ respectively; the effects of different parameters are illustrated in \Cref{fig:demo}.
The magnitude of each basis function in the fit is controlled by an amplitude parameter $a$.

\begin{table}
    \centering
    \caption{Priors on basis function parameters used in this paper.
Sorted priors have ordering enforced; see \citet[][Appendix A2]{Handley2015b} for more details.
The half Gaussian prior on the amplitudes of the tanh basis functions is truncated at zero and permits only positive values.}%
    \label{tab:priors}
    \begin{tabular}{lll}
    \toprule
    Parameter         & Prior Type          & Prior Parameters                \\
    \midrule                      
    \multicolumn{3}{l}{1-dimensional generalised Gaussian~\eqref{equ:gg_1d}}     \\
    $N$               & Uniform (integer)   & $\in \mathbb{Z} \cap [1, 5]      $ \\
    $a$               & Sorted Exponential  & $\lambda=1                       $ \\
    $\mu$             & Uniform             & $\in [0, 1]                      $ \\
    $\sigma$          & Uniform             & $\in [0.03, 1.0]                 $ \\
    $\beta$           & Exponential         & $\lambda=0.5                     $ \\
    \midrule                      
    \multicolumn{3}{l}{1-dimensional tanh~\eqref{equ:ta_1d}}                     \\
    $N$               & Uniform (integer)   & $\in \mathbb{Z} \cap [1, 5]      $ \\
    $a$               & Sorted Half Gaussian& $\mu=0,\sigma=5                  $ \\
    $w$               & Gaussian            & $\mu=0,\sigma=5                  $ \\
    $b$               & Gaussian            & $\mu=0,\sigma=5                  $ \\
    \midrule                      
    \multicolumn{3}{l}{Adaptive basis function family selection}                 \\
    $T$               & Uniform (integer)   & $\in \mathbb{Z} \cap [1, 2]      $ \\
    \midrule                      
    \multicolumn{3}{l}{2-dimensional generalised Gaussian~\eqref{equ:gg_2d}}     \\
    $N$               & Uniform (integer)   & $\in \mathbb{Z} \cap [1, 5]      $ \\
    $a$               & Sorted Exponential  & $\lambda=1                       $ \\
    $\mu_1$           & Uniform             & $\in[0, 1]                       $ \\
    $\mu_2$           & Uniform             & $\in[0, 1]                       $ \\
    $\sigma_1$        & Uniform             & $\in [0.03, 0.5]                 $ \\
    $\sigma_2$        & Uniform             & $\in [0.03, 0.5]                 $ \\
    $\beta_1$         & Exponential         & $\lambda=0.5                     $ \\
    $\beta_2$         & Exponential         & $\lambda=0.5                     $ \\
    $\Omega$          & Uniform             & $\in [-\pi/4,\pi/4]              $ \\
    \bottomrule
    \end{tabular}
\end{table}

When $\beta=2$,~\eqref{equ:gg_1d} is proportional to a normal distribution with variance $\sigma^2 / 2$, and when $\beta = 1$ it is proportional to a Laplace distribution.
For large values of $\beta$,~\eqref{equ:gg_1d} is approximately uniform $\in [\mu - \sigma, \mu + \sigma]$ and zero elsewhere.
The normalisation constant $\beta / (\Gamma(\frac{1}{\beta})2\sigma)$ is omitted from~\eqref{equ:gg_1d} as it causes pronounced degeneracies in the joint posterior distributions of $a$, $\beta$ and $\sigma$ due to all 3 parameters affecting the height of the basis function at its centre.

The priors used for the basis functions are shown in \Cref{tab:priors}.
We use a uniform prior on the number and type of basis functions, but other priors can be used when they are justified for the problem considered --- for example a prior favouring small values of $N$ will result in more sparse solutions.
The exponential prior on the amplitudes $\bm{a}$ of the generalised Gaussians has the desirable property that it is a function of only the sum of the amplitudes and does not vary based on how the total is split between basis functions.
We use a uniform prior on the generalised Gaussians' $\sigma$, rather than a scale prior favouring smaller values, as we find the latter causes overfitting by encouraging the addition of narrow generalised Gaussians to fit noise in the data.
The priors on the tanh basis functions are chosen for consistency with the neural networks discussed in \Cref{sec:nn}.

\subsection{Numerical results}

\begin{figure*}
	\centering
    \begin{subfigure}{\linewidth}
        \includegraphics{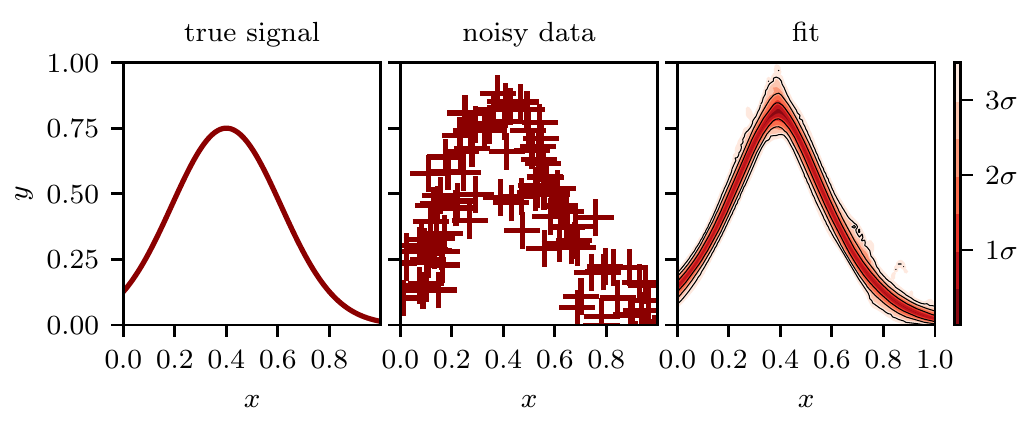}
        \includegraphics{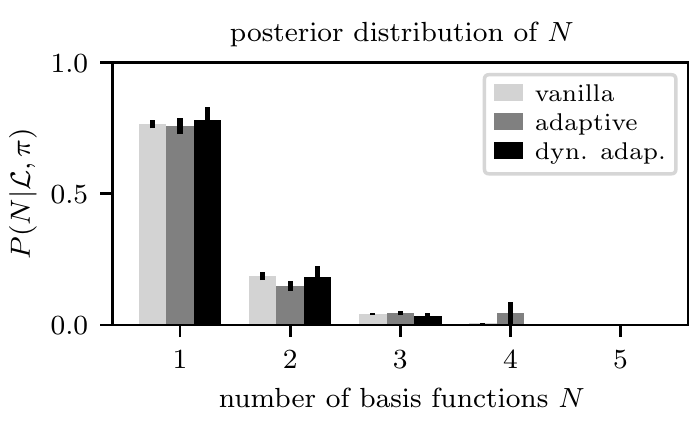}
        \subcaption{Data from a single generalised Gaussian.}%
        \label{sub:multi_gg_1d_gg_1d_1}
    \end{subfigure}
    \begin{subfigure}{\linewidth}
        \includegraphics{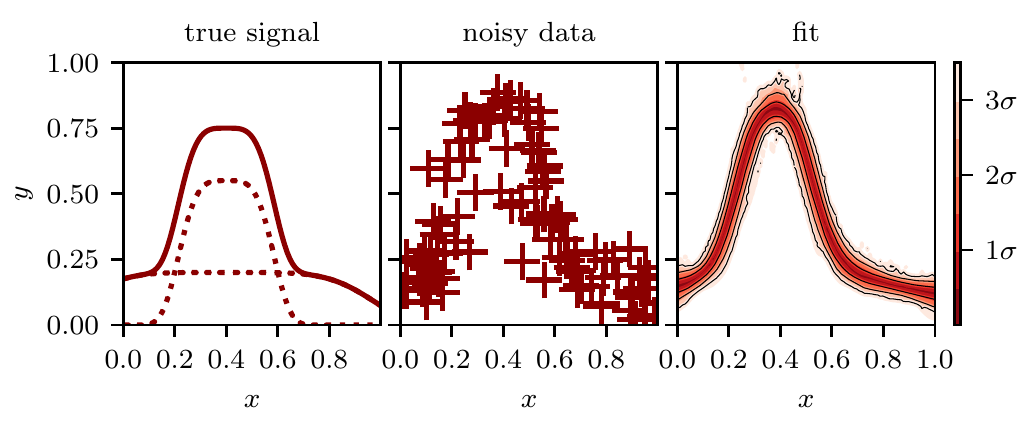}
        \includegraphics{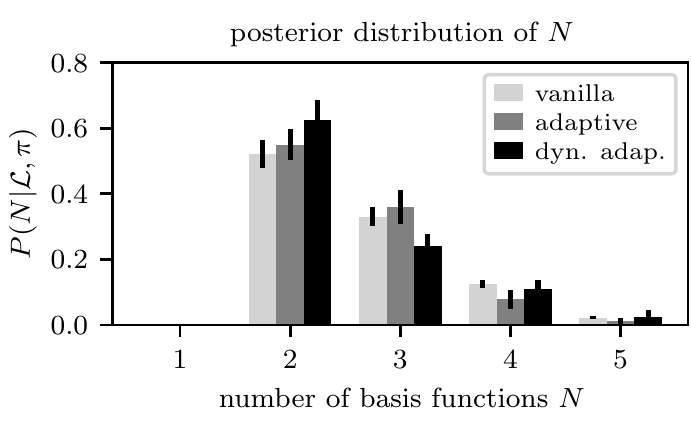}
        \subcaption{Data from the sum of two generalised Gaussians.}%
        \label{sub:multi_gg_1d_gg_1d_2}
    \end{subfigure}
    \begin{subfigure}{\linewidth}
        \includegraphics{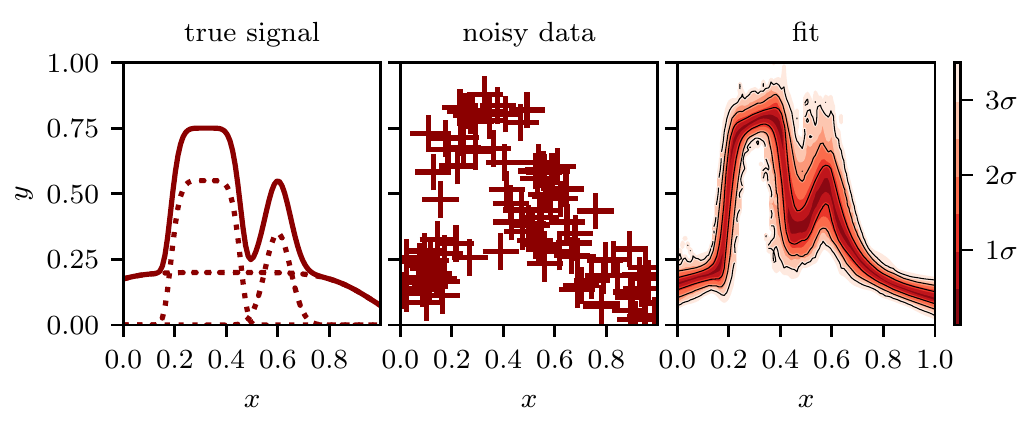}
        \includegraphics{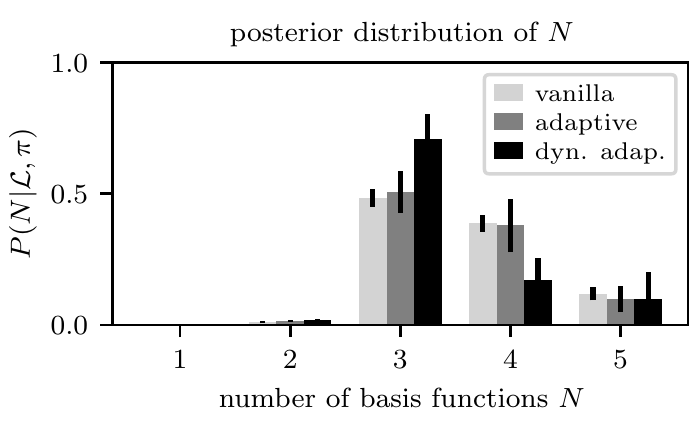}
        \caption{Data from the sum of three generalised Gaussians.}%
        \label{sub:multi_gg_1d_gg_1d_3}
    \end{subfigure}
    \caption{Fitting generalised Gaussian basis functions to 100 data points sampled from different combinations of basis functions.
    In each row the first plot shows the true signal (a sum of basis functions); where this contains more than one basis function, the individual components are shown with dashed lines.
	The data, which includes added normally distributed $x$- and $y$-errors with $\sigma_x=\sigma_y=0.07$, is show in the second plot.
The third plot shows the fit calculated using the adaptive method with dynamic nested sampling; coloured contours represent posterior iso-probability credible intervals on $y(x)$.
    The bar plots on the right display the posterior distribution for different numbers of basis functions $N$; values calculated using the vanilla method and using the adaptive method with standard nested sampling are also included for comparison.
    Results shown for the adaptive method use a combined inference from 5 runs, each of which computes a full posterior on $N$ and uses 1,000 live points; adaptive runs using dynamic nested sampling have \dyPolyChord{} settings $n_\mathrm{init}=500$ and $\texttt{dynamic\_goal}=1$.
    Results for the vanilla method use 5 separate runs, each with 200 live points, to compute the evidence for each value of $N$.
    All runs use the setting $\texttt{num\_repeats}=100$.
    The parameters of the basis functions in the true signal and numerical results for the computational efficiency of the different methods are shown in \Cref{tab:gg_1d_data_args} and \Cref{tab:gg_1d_gg_1d_results} respectively in Appendix~\ref{app:anr}.}%
    \label{fig:multi_gg_1d_gg_1d}
	\centering
    \includegraphics{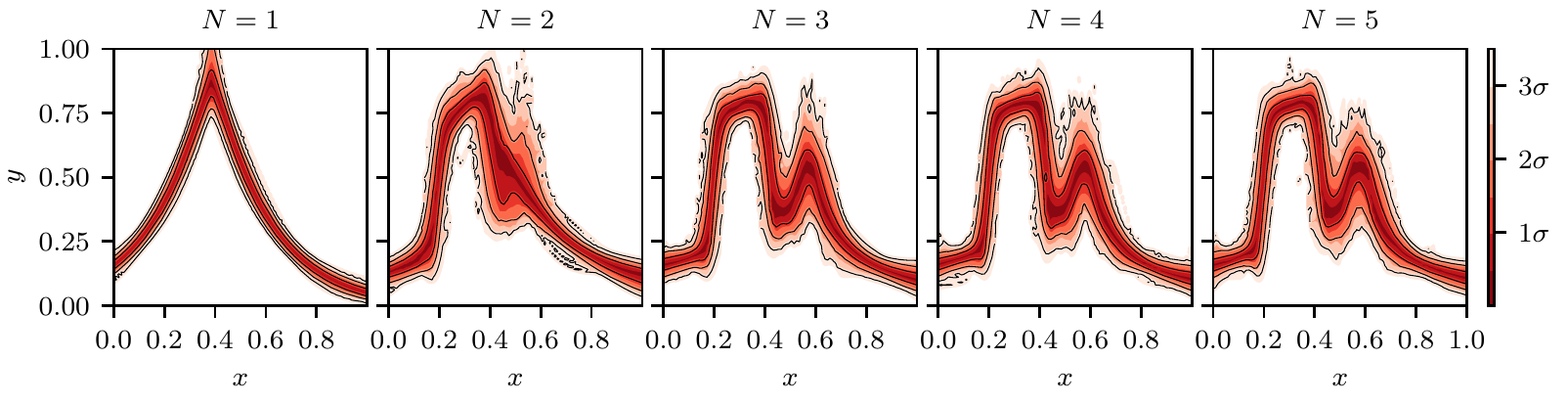}
    \caption{Fits of the data shown in \Cref{sub:multi_gg_1d_gg_1d_3} conditioned on different numbers $N$ of basis functions.
	These plots are made using the vanilla method nested sampling runs, as the adaptive method runs contain relatively few samples from the heavily disfavoured values of $N$.}%
    \label{fig:split_gg_1d_gg_1d}
\end{figure*}
\begin{figure*}
    \vspace{1cm}  
	\centering
    \begin{subfigure}{\linewidth}
        \includegraphics{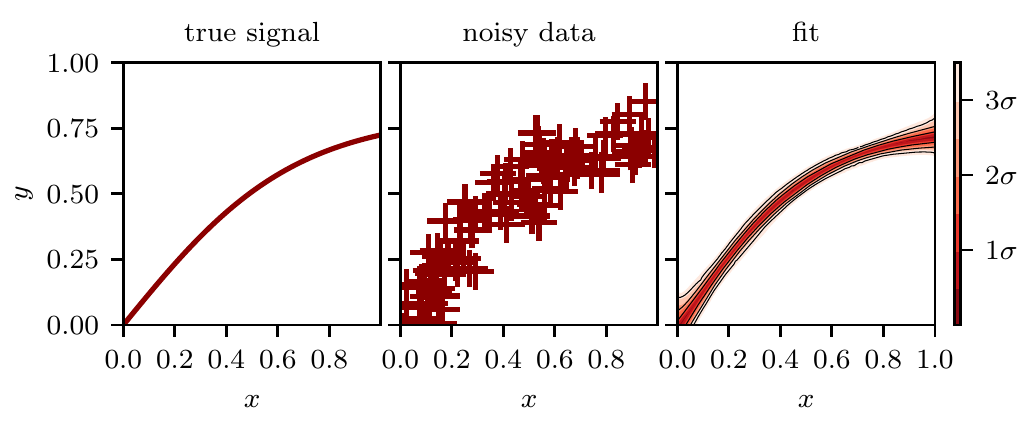}
        \includegraphics{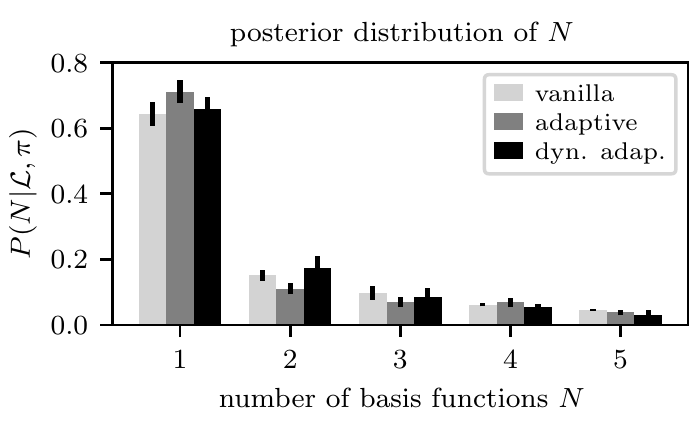}
        \subcaption{Data from a single tanh basis function.}%
        \label{sub:multi_ta_1d_ta_1d_1}
    \end{subfigure}
    \begin{subfigure}{\linewidth}
        \includegraphics{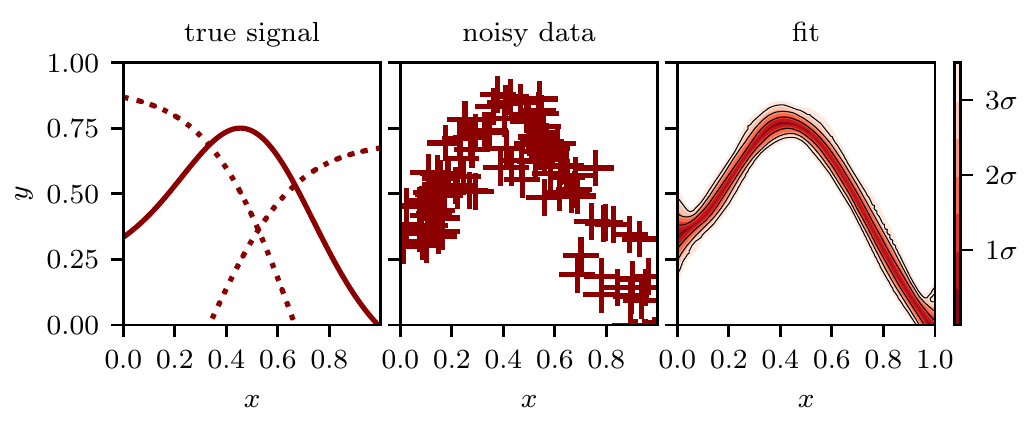}
        \includegraphics{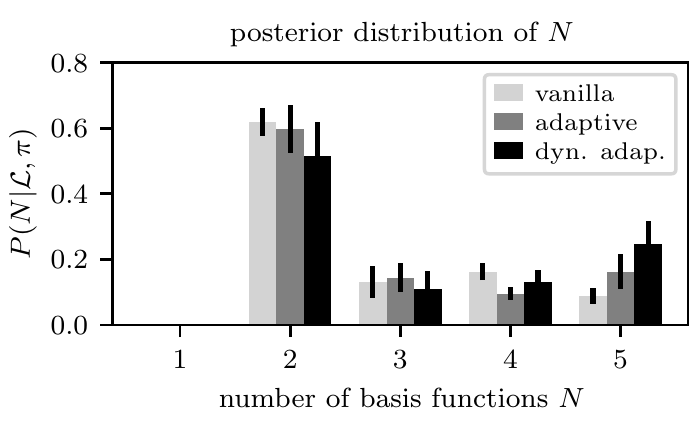}
        \subcaption{Data from the sum of two tanh basis functions.}%
        \label{sub:multi_ta_1d_ta_1d_2}
    \end{subfigure}
    \begin{subfigure}{\linewidth}
        \includegraphics{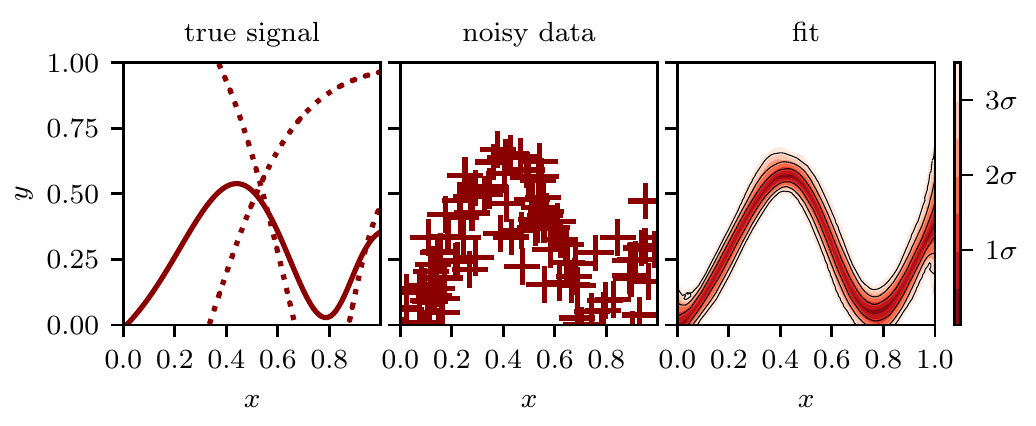}
        \includegraphics{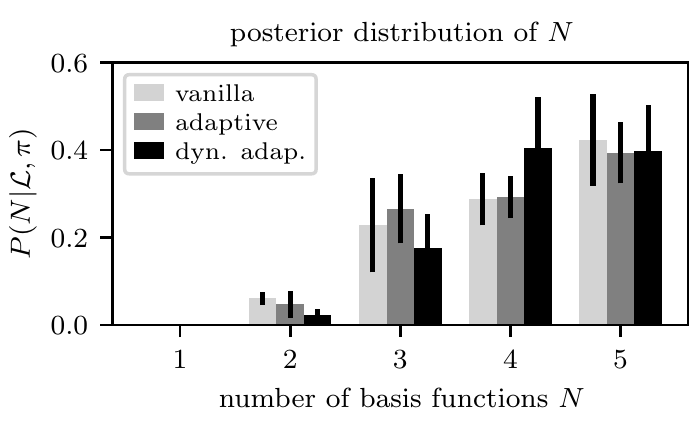}
        \caption{Data from the sum of three tanh basis functions.}%
        \label{sub:multi_ta_1d_ta_1d_3}
    \end{subfigure}
    \caption{As for \Cref{fig:multi_gg_1d_gg_1d} but using tanh basis functions instead of generalised Gaussians.
    The parameters of the tanh basis functions in true signal are shown in \Cref{tab:ta_1d_data_args} in Appendix~\ref{app:anr}.}%
    \label{fig:multi_ta_1d_ta_1d}
    \includegraphics{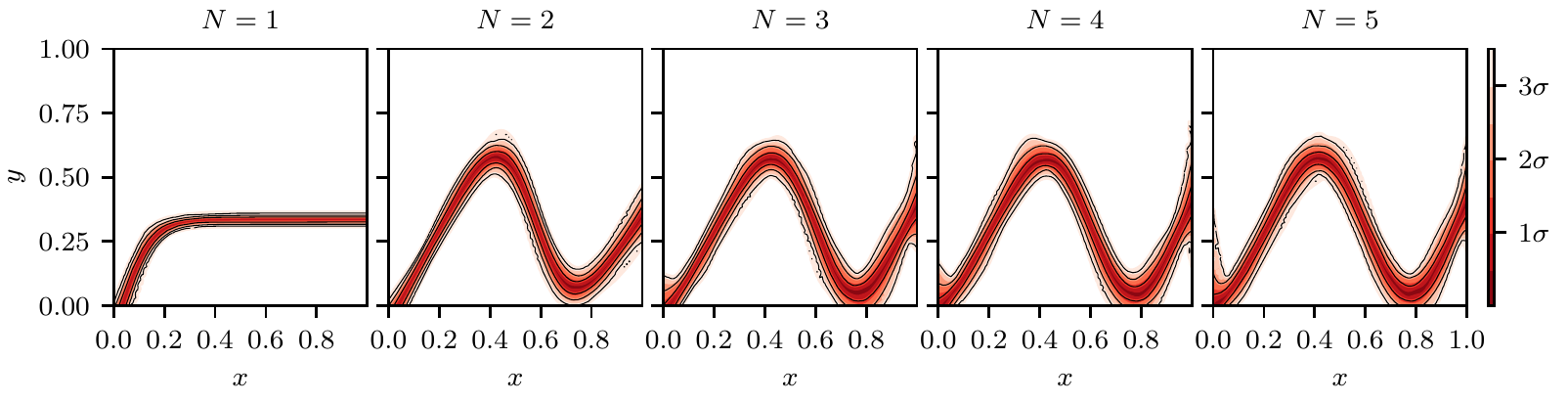}
    \caption{Fits of the data sets shown in \Cref{sub:multi_ta_1d_ta_1d_3} conditioned on different numbers $N$ of basis functions.}%
    \label{fig:split_ta_1d_ta_1d}
    \vspace{1cm}  
\end{figure*}
\begin{figure*}
	\centering
    \begin{subfigure}{\linewidth}
        \includegraphics{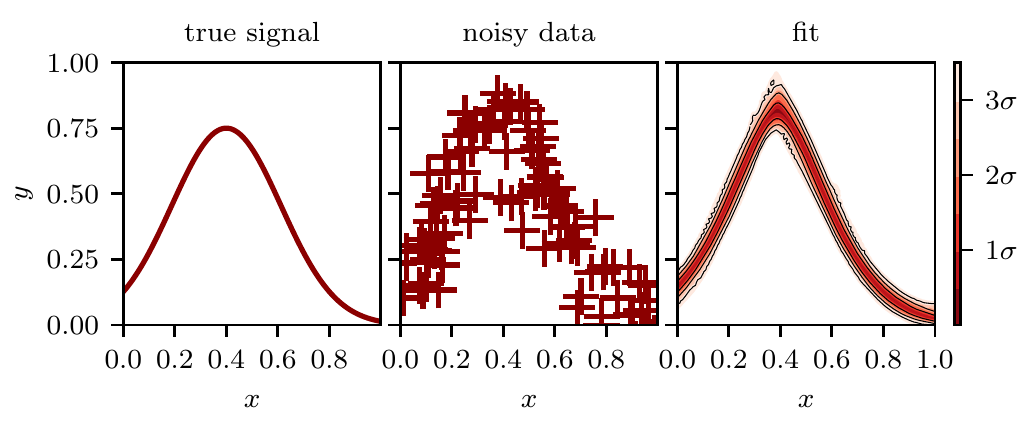}
        \includegraphics{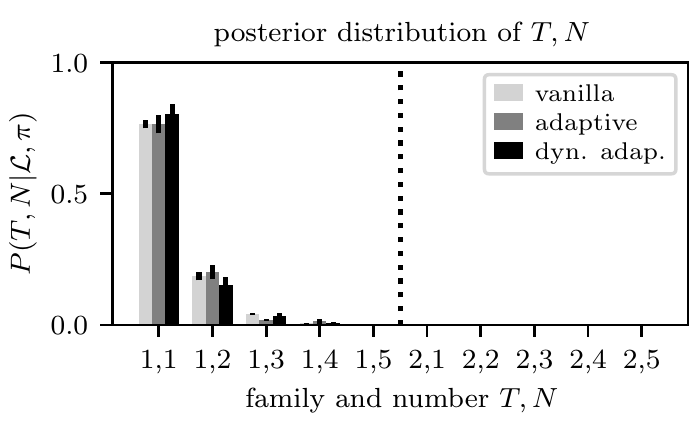}
        \subcaption{Data used in \Cref{sub:multi_gg_1d_gg_1d_1} from a single generalised Gaussian.}%
        \label{sub:multi_adfam_gg_ta_1d_gg_1d_1}
    \end{subfigure}
    \begin{subfigure}{\linewidth}
        \includegraphics{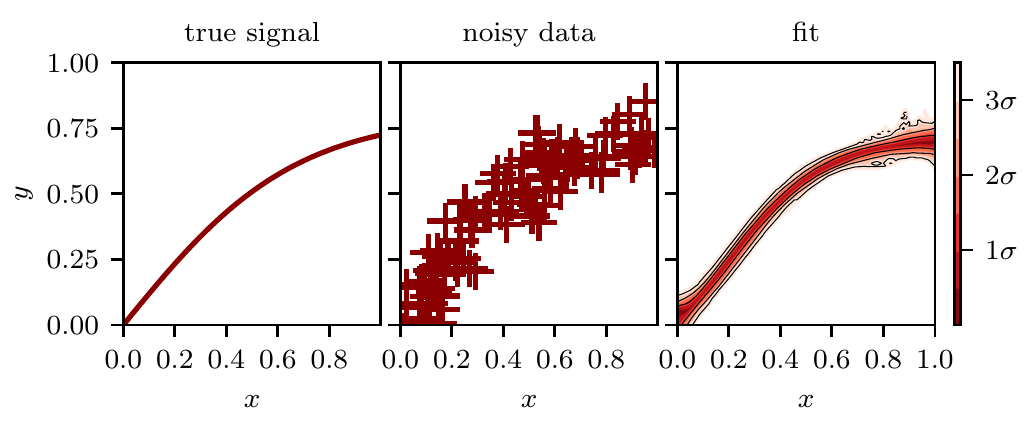}
        \includegraphics{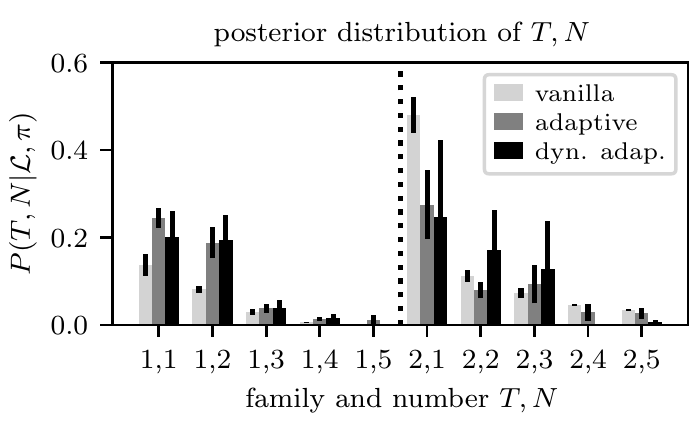}
        \subcaption{Data used in \Cref{sub:multi_ta_1d_ta_1d_1} from a single tanh function.}%
        \label{sub:multi_adfam_gg_ta_1d_ta_1d_1}
    \end{subfigure}
    \caption{Fitting generalised Gaussian and tanh basis functions to data in a fully adaptive manner with the family determined by an integer parameter $T$.
    In each row the first plot shows the true signal (a sum of basis functions).
	The data, shown in the second plot, contains normally distributed $x$- and $y$-errors with $\sigma_x=\sigma_y=0.07$.
    The third plot shows the fit calculated using the adaptive method with dynamic nested sampling; coloured contours represent posterior iso-probability credible intervals on $y(x)$.
    The bar plots on the right display the posterior posterior distribution on different families $T$ and numbers of basis functions $N$; values calculated using the vanilla method and using the adaptive method with standard nested sampling are also included for comparison.
    Results for the adaptive method use a combined inference from 5 runs, each of which computes a full posterior on $T,N$ and uses 1,000 live points; adaptive runs using dynamic nested sampling have \dyPolyChord{} settings $n_\mathrm{init}=500$ and $\texttt{dynamic\_goal}=1$.
    Results for the vanilla method use separate runs, each with 200 live points, to compute the evidence for each combination of $T$ and $N$.
    All runs use the setting $\texttt{num\_repeats}=100$.}%
    \label{fig:multi_adfam_gg_ta_1d}
\end{figure*}

We first illustrate Bayesian sparse reconstruction using simulated 1-dimensional data points sampled from basis function mixture models.
Independent Gaussian $x$- and $y$-errors of size $\sigma_x=\sigma_y=0.07$ are added to each data point.

\Cref{fig:multi_gg_1d_gg_1d} show the results from fitting different Gaussian mixture models; plots of the posterior distribution of $y$ were created using the \fgivenx{} package \citep{Handley2018fgivenx}.
Despite the large measurement noise and visually similar data in \Cref{sub:multi_gg_1d_gg_1d_1,sub:multi_gg_1d_gg_1d_2,sub:multi_gg_1d_gg_1d_3}, our approach is able to correctly reconstruct the different true signals and identify the increasing number of basis functions required to model each signal.

\Cref{fig:split_gg_1d_gg_1d} shows examples of signal reconstructions conditioned on specific numbers of basis functions, and illustrates the effect of increasing $N$ on the fit produced.
Such plots can be calculated from adaptive method nested sampling runs by marginalising over different values for the integer parameter $N$.
However, we use the vanilla method runs to make these plots as the adaptive method dedicates relatively few samples to exploring the highly disfavoured $N$ values with make negligible contribution to the overall fit.
This is a desirable feature which makes fitting with the adaptive method more efficient, but as a consequence the vanilla method can produce more accurate plots conditioned on disfavoured values of $N$.

\Cref{fig:multi_ta_1d_ta_1d,fig:split_ta_1d_ta_1d} show examples of fitting tanh basis functions to 1-dimensional data, and are similar to~\Cref{fig:multi_gg_1d_gg_1d,fig:split_gg_1d_gg_1d}.
As for the generalised Gaussian basis functions, our approach is able to accurately reconstruct the true signal from the noisy data and identify the increasing complexity of the successive signals in \Cref{sub:multi_ta_1d_ta_1d_1,sub:multi_ta_1d_ta_1d_2,sub:multi_ta_1d_ta_1d_3}.
However it is not necessarily the case that the most probable {\em a posteriori\/} value of $N$, given the noisy data and priors, is the same as the number of basis functions from which the data was sampled; in \Cref{sub:multi_ta_1d_ta_1d_3}, $P(N=4|\mathcal{L},\pi)$ and $P(N=5|\mathcal{L},\pi)$ are greater than $P(N=3|\mathcal{L},\pi)$.

\subsection{Adaptive basis function families}\label{sec:adfam}

We now illustrate including a second integer parameter $T$ which selects the basis function family, in addition to $N$ which selects the number of basis functions given the family.
\Cref{fig:multi_adfam_gg_ta_1d} shows fits using both generalised Gaussians ($T=1$) and tanh functions ($T=2$), with a uniform prior on $T \in \mathbb{Z} \cap [1, 2]$.

The generalised Gaussians are a much better fit for the data in \Cref{sub:multi_adfam_gg_ta_1d_gg_1d_1}, with posterior probability from the adaptive method with dynamic nested sampling of $P(T=2|\mathcal{L},\pi) = (1\pm1)\times 10^{-7}$.
In contrast the two families are competitive for the data in \Cref{sub:multi_adfam_gg_ta_1d_ta_1d_1}, with $P(T=2|\mathcal{L},\pi) = 0.6\pm0.1$ indicating only a weak favouring of the tanh basis function.
These results are, however, highly dependent on the priors used for the basis functions' parameters.

A possible application of this adaptive selection of basis function families $T$ would be to compare different parametric models for sources in astronomical images in which the true number of sources is unknown.
In this case computing a posterior distribution on $T$ would not only marginalise over the distributions of the sources' parameters, but also over the unknown number of sources $N$.

\subsection{Comparison of vanilla and adaptive results}\label{sec:comp_vanilla_adaptive}

The adaptive method allows significant improvements in accuracy of the overall fit for a given computational cost by allocating fewer samples into disfavoured models which make a small or negligible contribution to the output.
In addition, by transforming the model selection from evidence calculations (as in the vanilla method) to a parameter estimation problem on $N$, the adaptive method changes the nature of the sampling errors.
Uncertainty in the rate of shrinkage at each step before any significant posterior mass is reached --- the dominant source of error in nested sampling evidence calculations --- has a negligible effect on parameter estimation of the posterior distribution of $N$.
This can allow order-of-magnitude gains in computational efficiency of posterior odds ratios from the adaptive method compared to the vanilla method, as observed by \citet{Chua2018}.
However, a downside of the method is that including all the models and the integer parameter makes the posterior distribution highly multimodal and more challenging for the sampler to explore.

The errors on nested sampling calculations scale in inverse proportion to the square root of the computational effort used, and for a given likelihood and prior the number of samples produced is roughly proportional to the computational effort.
Following \citet{Higson2017b} we therefore measure the increase in computational efficiency (adjusted for any differences in the number of samples taken) from alternative methods compared to the vanilla method with standard nested sampling as
\begin{equation}
    \mathrm{efficiency\,gain} = \frac{\mathrm{Var}[\mathrm{vanilla\,NS\,results}]}{\mathrm{Var}[\mathrm{method\,NS\,results}]} \times \frac{\overline{N_{\mathrm{samp,van}}}}{\overline{N_{\mathrm{samp,meth}}}}.
    \label{equ:efficiency_gain_bsr}
\end{equation}
Here the first term is the ratio of the estimated variance of the results of repeated calculations using the vanilla method and the alternative method; the second term is the ratio of the mean number of samples from the nested sampling runs using each method.
Numerical results for the efficiency gains from the different methods are show in \Cref{tab:gg_1d_gg_1d_results} in Appendix~\ref{app:anr}.
These use estimates of the variance of results calculated using the bootstrap resampling method described in \citet{Higson2017a}, which avoids the need to compute large numbers of nested sampling runs but also does not include additional errors from the sampler failing to explore the parameter space fully \citep[see][for a detailed discussion of such errors]{Higson2018a}.
As described in Appendix~\ref{app:anr}, we find that the sampler is not able to explore the parameter space perfectly with the settings used, meaning the true variance of results is higher than the bootstrap estimates.
As a result, given the adaptive method's more complex posterior distribution, the efficiency gains of factors of up to $14\pm3$ for the adaptive method and $46\pm 9$ for the adaptive method using dynamic nested sampling are likely to be overestimates and are best viewed as an indication of what is possible using the method with more computational power (such as using a higher value for \dyPolyChord's \numrepeats{} setting).

\section{2-dimensional image fitting}\label{sec:2d_bsr}

We now demonstrate Bayesian sparse reconstruction for monochrome images.
Here, for each data point (pixel) $d$, $\bm{x}_d={(x_{1},x_{2})}_d$ is the pixel location and $y_d \in [0,1]$ is the scalar signal.
For simplicity we assume that the errors in the pixel positions $\bm{x}_d$ are negligible, and consider the case that the signal $y_d$ for each pixel contains independent Gaussian noise with size $\sigma_y = 0.2$ --- in this case the likelihood is given by~\eqref{equ:fitting_likelihood}.

We define 2-dimensional generalised Gaussians as the product of two 1-dimensional generalised Gaussians~\eqref{equ:gg_1d} rotated by angle $\Omega$ around their mean $\bm{\mu}$:
\begin{equation}
    \begin{aligned}
        \phi^{(\mathrm{g2d})}(\bm{a},\bm{p})
        &=
        \phi^{(\mathrm{g2d})}(\bm{x},\bm{\mu},\bm{\sigma},\bm{\beta},\Omega) \\
        &=
        \phi^{(\mathrm{g1d})}(x'_1,\mu_1,\sigma_1,\beta_1) \times \phi^{(\mathrm{g1d})}(x'_2,\mu_2,\sigma_2,\beta_2) \\
    \mathrm{where} \quad \bm{x'} &= \bm{\mu} + (\bm{x} - \bm{\mu})
     \begin{pmatrix}
    \cos(\Omega) & -\sin(\Omega)  \\
    \sin(\Omega) &\cos(\Omega)  
  \end{pmatrix}.
    \label{equ:gg_2d}
    \end{aligned}
\end{equation}
The priors used are shown in \Cref{tab:priors}. 

\begin{figure*}
	\centering
    \begin{subfigure}{\linewidth}
        \includegraphics{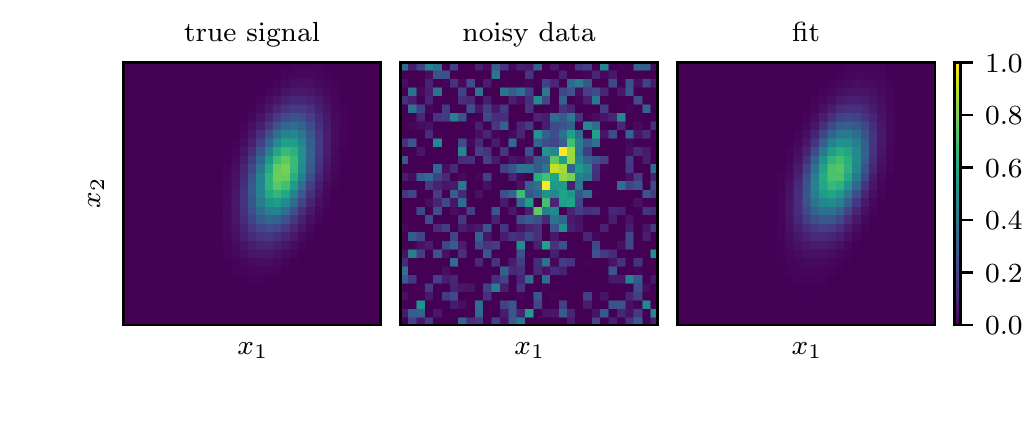}
        \includegraphics{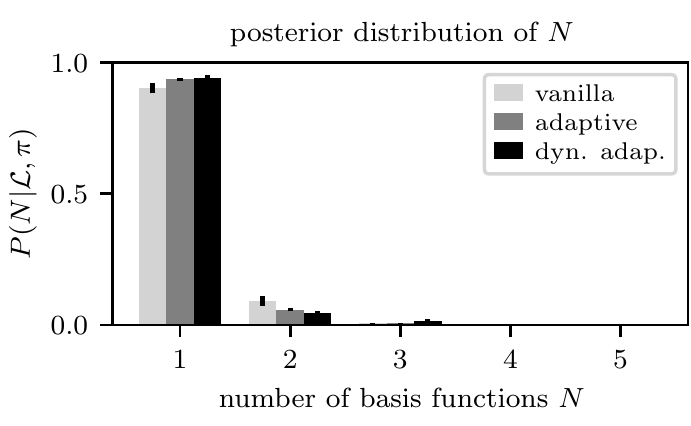}
        \subcaption{Image of a single 2-dimensional generalised Gaussian.}%
        \label{sub:multi_gg_2d_gg_2d_1}
    \end{subfigure}
    \begin{subfigure}{\linewidth}
        \includegraphics{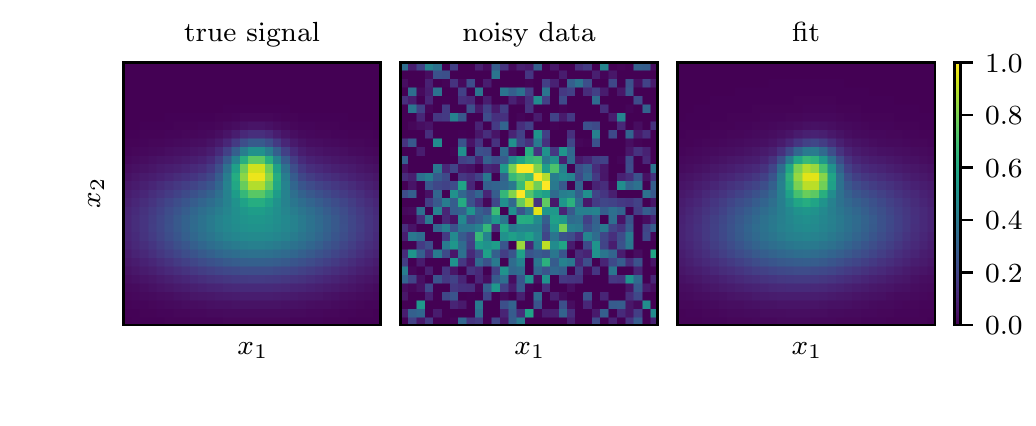}
        \includegraphics{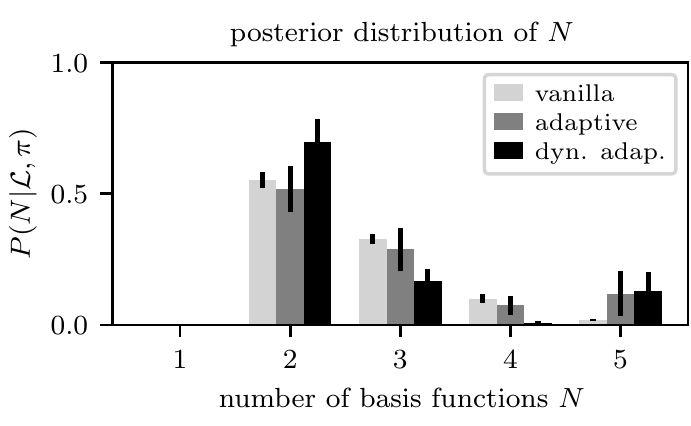}
        \subcaption{Image of the sum of two 2-dimensional generalised Gaussians.}%
        \label{sub:multi_gg_2d_gg_2d_2}
    \end{subfigure}
    \begin{subfigure}{\linewidth}
        \includegraphics{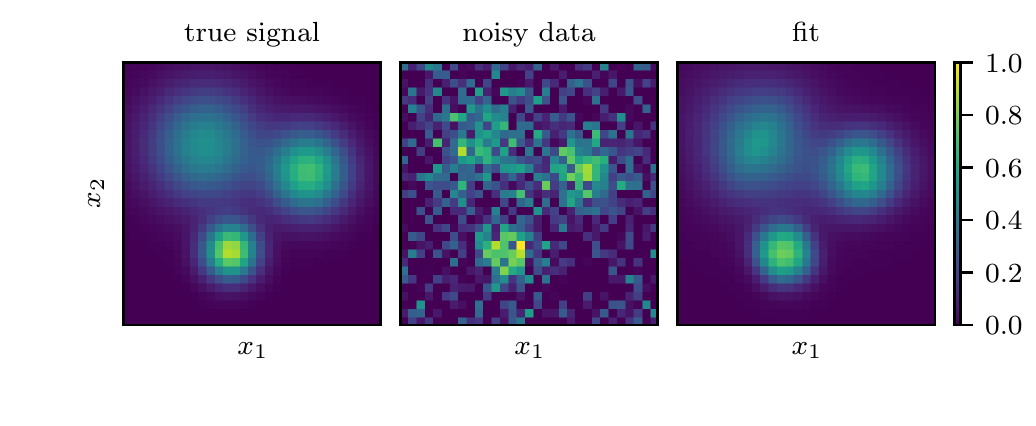}
        \includegraphics{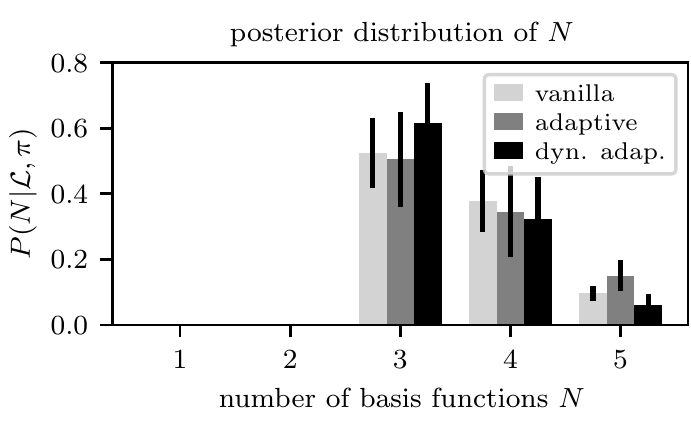}
        \caption{Image of the sum of three 2-dimensional generalised Gaussians.}%
        \label{sub:multi_gg_2d_gg_2d_3}
    \end{subfigure}
    \caption{Fitting 2-dimensional generalised Gaussian basis functions to $32\times32$ images of mixtures of generalised Gaussians.
    In each row the 2 plots on the left show the true signal and the data, which has added normally distributed $y$-errors with $\sigma_y=0.2$.
    The third column shows the mean value of $y(\bm{x})$ from the posterior samples produced using the adaptive method with dynamic nested sampling.
    The bar plots on the right display the posterior distribution for different numbers of basis functions $N$; values calculated using the vanilla method and adaptive method without dynamic nested sampling are also included for comparison.
    Results for the adaptive method show a combined inference from 5 runs, each of which computes a full posterior on $N$ and uses 2,000 live points; adaptive runs using dynamic nested sampling have \dyPolyChord{} settings $n_\mathrm{init}=1,000$ and $\texttt{dynamic\_goal}=1$.
    Results for the vanilla method use 5 separate runs, each with 400 live points, to compute the evidence for each value of $N$.
    All runs use the setting $\texttt{num\_repeats}=250$.
    The parameters of the basis functions in the true signal and numerical results for the computational efficiency of the different methods are shown in \Cref{tab:gg_2d_data_args} and \Cref{tab:gg_2d_gg_2d_results} respectively in Appendix~\ref{app:anr}.}%
    \label{fig:multi_gg_2d_gg_2d}
    \includegraphics{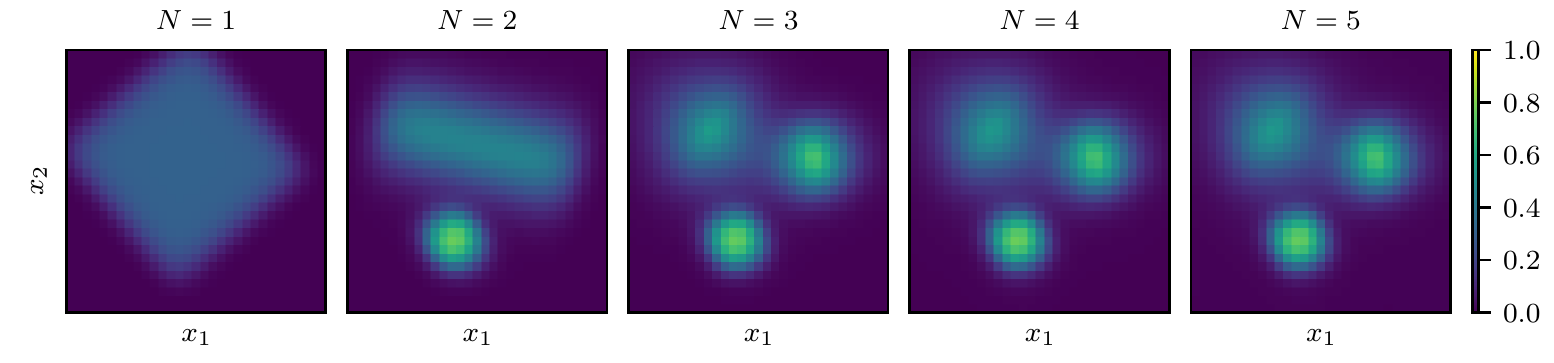}
    \caption{Fits of the data in \Cref{sub:multi_gg_2d_gg_2d_3} conditioned on different numbers $N$ of basis functions; these plots use results from the vanilla method.}%
    \label{fig:split_gg_2d_gg_2d}
\end{figure*}

\Cref{fig:multi_gg_2d_gg_2d} show examples of Bayesian sparse reconstruction fitting 2-dimensional images.
The fits show the mean values predicted for each pixel, averaged over all the samples produced in proportion to their posterior weight.
Using the mean value avoids overfitting --- which would occur if, for example, the fit was simply calculated from the sample with the highest likelihood (the maximum likelihood estimate).
The samples provide a full posterior distribution on the parameters and output signal, so other quantities such as the uncertainty on each pixel can also be easily calculated.
\Cref{fig:split_gg_2d_gg_2d} shows fits conditioned on specific values of $N$, and illustrates how increasing the number of basis functions allows increasingly complex structure to be included in the recovered image.

As in the 1-dimensional case, our approach is able to faithfully reconstruct the signal from the noisy data and the numbers of basis functions with the highest posterior probability (shown in the bar charts on the left of each subfigure) match the number of components in the mixture model used for the signal.
Furthermore, \Cref{tab:gg_2d_gg_2d_results} in Appendix~\ref{app:anr} shows efficiency gains~\eqref{equ:efficiency_gain_bsr} from the adaptive method of up to $10\pm2$ and from the adaptive method with dynamic nested sampling of up to $16\pm3$ in these cases.
However, as discussed in \Cref{sec:comp_vanilla_adaptive}, these numbers may overestimate the efficiency gains observed in practice with the settings used.

\subsection{Application to astronomical images}\label{sec:gg_2d_get_image}

We now apply the 2-dimensional fitting techniques from the previous section to astronomical images from the Hubble Space Telescope eXtreme Deep Field \citep{Illingworth2013}.
These are not ``true signals'' as in the previous examples because the images contain some measurement uncertainty, but this is relatively small compared to our added Gaussian errors of $\sigma_y=0.2$.
We therefore use them as an approximation of a realistic physical signal for testing our method.
Furthermore, for this first trial application of our method, we provide only a visual demonstration of the accuracy of our image reconstructions (to be assessed qualitatively).
A more quantitative evaluation can be performed in the future using simulations where the noise-free signal values are available.

\begin{figure*}
    \vspace{1cm}  
	\centering
    \begin{subfigure}{\linewidth}
        \includegraphics{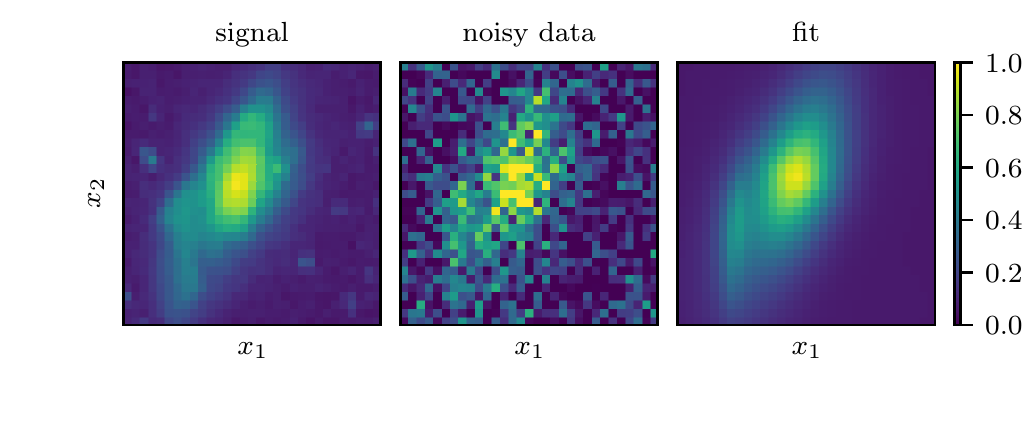}
        \includegraphics{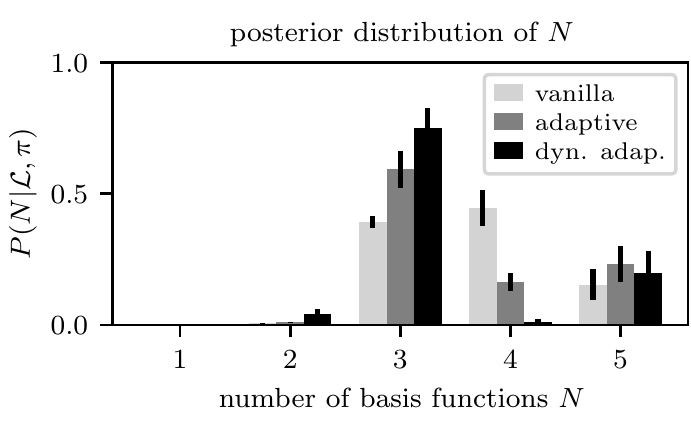}
        \subcaption{Image of an irregularly shaped galaxy.}%
        \label{sub:multi_gg_2d_get_image_1}
    \end{subfigure}
    \begin{subfigure}{\linewidth}
        \includegraphics{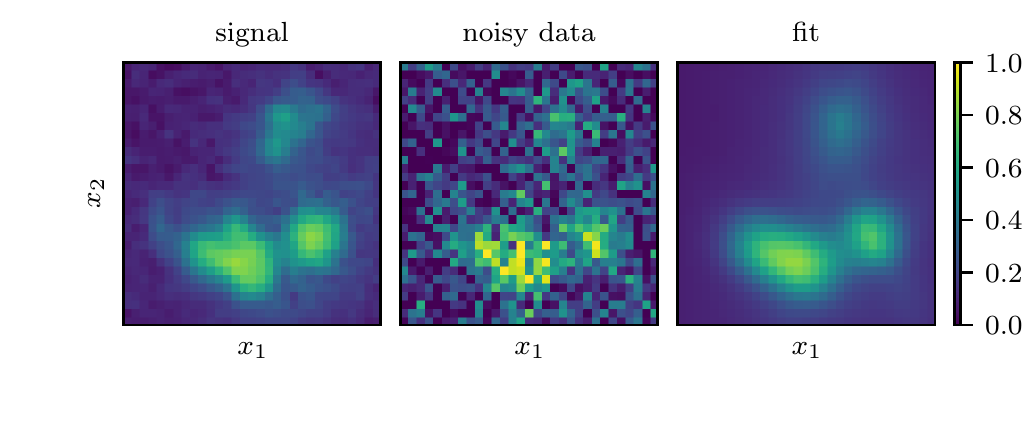}
        \includegraphics{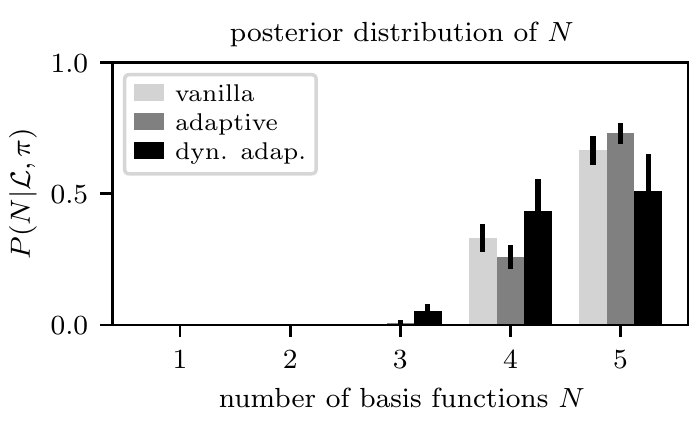}
        \subcaption{Image containing several galaxies.}%
        \label{sub:multi_gg_2d_get_image_2}
    \end{subfigure}
    \begin{subfigure}{\linewidth}
        \includegraphics{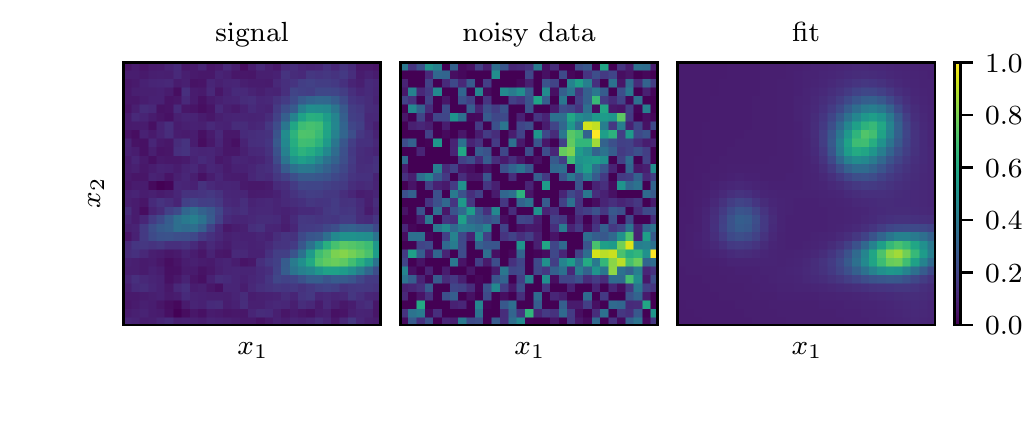}
        \includegraphics{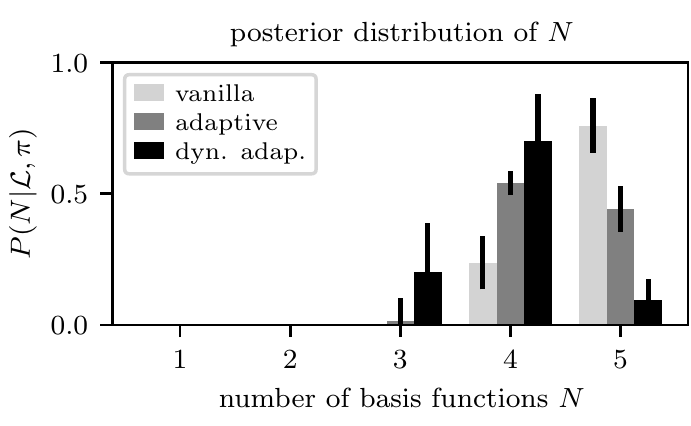}
        \caption{Another image containing several galaxies.}%
        \label{sub:multi_gg_2d_get_image_3}
    \end{subfigure}
    \caption{As for \Cref{fig:multi_gg_2d_gg_2d} but fitting $32\times32$ images from the Hubble Space Telescope eXtreme Deep Field \citep{Illingworth2013}; each pixel has added normally distributed $y$-errors with $\sigma_y=0.2$.}%
    \label{fig:multi_gg_2d_get_image}
    \includegraphics{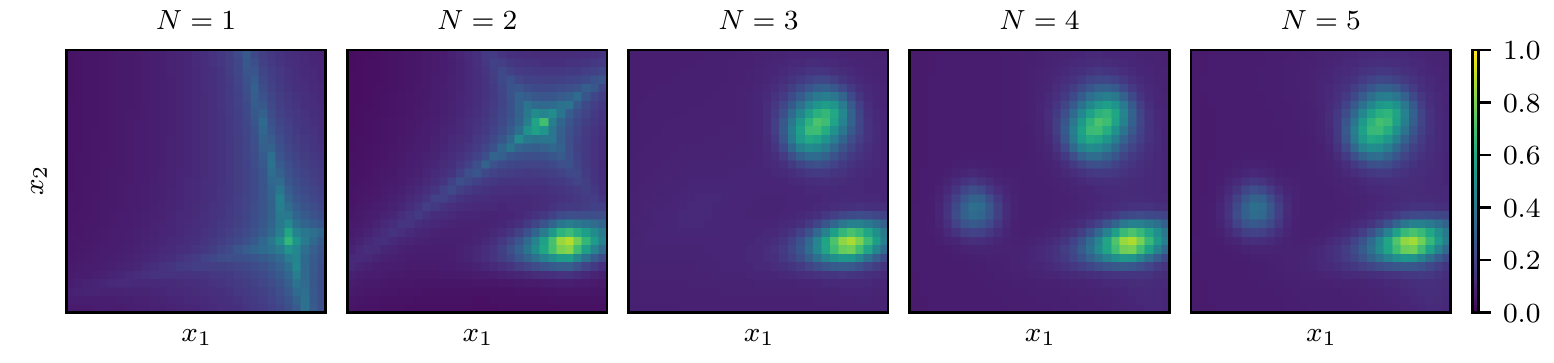}
    \caption{Fits of the data in \Cref{sub:multi_gg_2d_get_image_3} conditioned on different numbers of basis functions $N$; these plots use results from the vanilla method.}%
    \label{fig:split_gg_2d_get_image}
    \vspace{1cm}  
\end{figure*}

\Cref{fig:multi_gg_2d_get_image} shows fitting images of galaxies from the Hubble deep field using 2-dimensional generalised Gaussians~\eqref{equ:gg_2d}, and \Cref{fig:split_gg_2d_get_image} shows fits of specific numbers of basis functions (marginalised for different values of $N$).
Our method is able to faithfully reconstruct the signal from the noisy data, as can be seen from a visual comparison of the fit and the signal.
In this case, with the settings used, the posterior distributions of $N$ show some inconsistencies between the different methods.
These occur as in order to explore the challenging posterior consistently, \PolyChord{} and \dyPolyChord{} require higher live points and/or \numrepeats{} setting than those used; this leads to additional random errors.
However this lack of precision in the posterior probabilities of $N$ has little negative impact on the overall fit, as in each case all the posterior mass is allocated to values of $N$ which provide good representations of the data.

The posterior probabilities of different values of $N$ (shown on the left of each row of \Cref{fig:multi_gg_2d_get_image}) provide a measure of the complexity of the model justified by the data.
However, unless each basis function represents a justified physical model for the sources in the image, $N$ cannot necessarily be interpreted as the number of sources; for example a single source with a non-Gaussian structure may be represented by several Gaussian basis functions.

\section{Neural networks as adaptive basis regression}\label{sec:nn}

We now apply our Bayesian sparse reconstruction framework to artificial neural networks, where it allows a dynamic selection of the optimum network architecture.

\subsection{Background: feed forward neural networks}

Artificial neural networks (hereafter neural networks) are a popular machine learning technique loosely inspired by biological brains.
\citet[][Section V]{MacKay2003} provides a good introduction; for a detailed Bayesian reference see \citet{Neal2012b}.
Neural networks have been successfully applied to many areas of astronomical data analysis, including to image processing \citep[see for example][]{Graff2014,Ball2010}.

Neural networks are made up of nodes (``neurons'') which receive input signals and map them to a scalar signal (``activation''), which is then passed to other nodes.
We restrict our analysis to ``fully-connected'' ``feed-forward'' networks, in which nodes are arranged in layers and each node receive inputs from every node in the previous layer and passes its output to every node in the following layer (in this case the network is a directed acyclic graph).
Layers of nodes between the network's input and output are termed ``hidden layers'' as their outputs are not directly specified by the signal.

\begin{figure}
	\centering
    \includegraphics[width=0.8\linewidth]{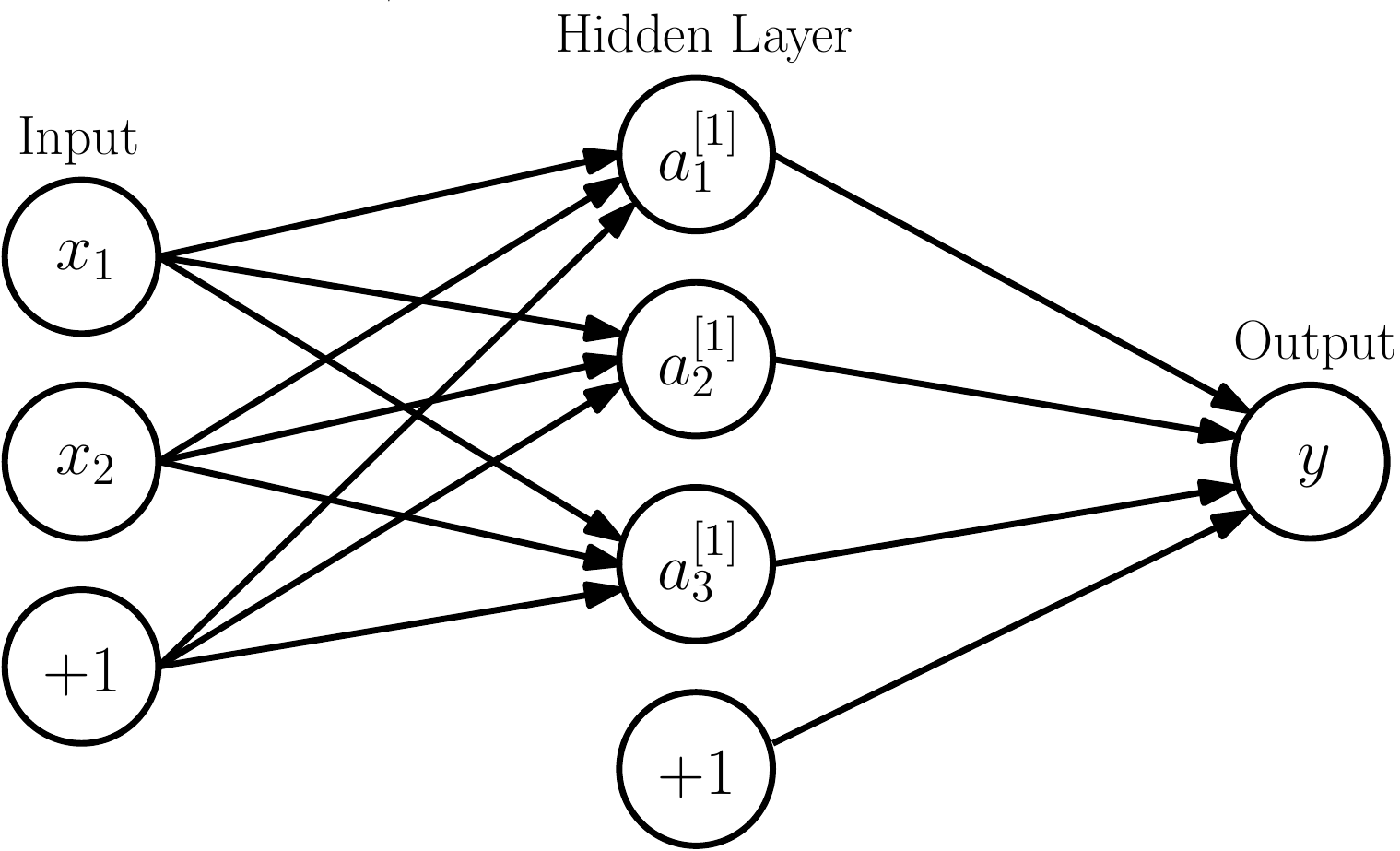}
    \caption{A feed-forward neural network with 2 inputs, a single hidden layer with 3 nodes and a scalar output $y$. Circles labelled $+1$ and the arrows leading from them represent the bias parameters $b_j^{[l]}$.}\label{fig:nn}
\end{figure}

Following the neural network literature, we denote the activation of the $j$\textsuperscript{th} node in the $l$\textsuperscript{th} layer as $a_j^{[l]}$; this is differentiated from the basis function amplitudes used earlier in the paper by the superscript label in square brackets and by the context.
The activation of each node is computed as
\begin{equation}
    a_j^{[l]} = \phi^{[l]}\left(\sum_{i=1}^{N^{[l-1]}} a^{[l-1]}_i w_{ji}^{[l]} + b_{j}^{[l]} \right),
\end{equation}
where $w_{j1}^{[l]},\dots,w_{jN^{[l-1]}}^{[l]}$ are the weights assigned to the activations of the $N^{[l-1]}$ nodes in the previous layer and conventionally an additional parameter $b_j^{[l]}$ (referred to as the ``bias'') is included.
The activation function $\phi^{[l]}$ is typically non-linear function of the inputs such as tanh or rectifier functions.\footnote{Rectifier functions such as $\phi(x) = \max(0, x)$ are now popular for deep neural networks, as they make it easier to optimise the network's weights with gradient-based methods because their gradient does not become small when $x$ is large \citep{Lecun2015}.
This paper uses the hyperbolic tangent function as we do not rely on gradient-based optimisation and our networks only have one or two hidden layers.}
For a feed-forward neural network with a $d$-dimension input $\bm{x}$ and one hidden layer containing $N$ nodes, the activations are:
\begin{eqnarray}
    a_j^{[1]}
    &=&
    \phi^{[1]}\left(\sum_{i=1}^d x_i w_{ji}^{[1]} + b_j^{[1]} \right)\label{equ:nn_1l_hidden},
    \\
    y_j
    &=& a_j^{[2]} =
    \phi^{[2]}\left(\sum_{i=1}^N a^{[1]}_i w_{ji}^{[2]} + b_j^{[2]} \right).\label{equ:nn_1l_output}
\end{eqnarray}
Such a network with a single output $y$ is illustrated in \Cref{fig:nn}.

Values for the network parameters can be selected using gradient-based optimisation and regularisation --- this is useful for ``deep learning'', in which networks have large numbers of hidden layers (are ``deep'') and sampling the posterior distribution over the full parameter space is not computationally feasible.
Often some part of the data set is held back and used for selecting the regularisation parameter.

Bayesian methods also provide a natural framework for neural networks, and simplified Bayesian computation can be performed in the space of neural network parameter values using techniques such as Bayes by backprop \citep{Blundell2015} and Gaussian approximations \citep{MacKay1995}.
In addition, many regularisation techniques commonly applied to neural networks can be interpreted from a Bayesian perspective \citep[see for example][]{Gal2016}.

\subsection{Applying Bayesian sparse reconstruction to neural networks}

To illustrate the connection between neural networks and the basis function fitting in previous sections of the paper, consider fitting a scalar signal $y$ using a signal hidden layer neural network in which the output layer has an identity activation function $\phi^{[2]}(x) = x$ and its bias parameter $b^{[2]}$ set to zero.
In this case the output~\eqref{equ:nn_1l_output} is simply a sum of basis functions.
If a tanh activation function is used for the nodes in the hidden layer and there is a scalar input $x$, such a network is equivalent to fitting 1-dimensional tanh basis functions as shown in~\Cref{fig:multi_ta_1d_ta_1d,fig:split_ta_1d_ta_1d}.
In this case, determining the value of $N$ using the framework introduced earlier in the paper represents Bayesian inference on the optimum number of nodes in the hidden layer given the data.

When there is more than one hidden layer, the output is no longer a direct sum of the inputs but our Bayesian sparse reconstruction framework can still be readily applied.
Furthermore the number of hidden layers $L$ can be determined by treating it as an integer parameter, in the same way the basis function family was represented by the integer parameter $T$ in \Cref{sec:adfam}.
We use the same number of nodes $N$ in each hidden layer, but if required one could allow the hidden layers to have different numbers of nodes governed by multiple integer parameters $N^{[1]}, \dots, N^{[L]}$.
We consider only a single output for simplicity, but our results easily generalises to neural networks with multiple outputs $(y_1, y_2, \dots)$ and to classification problems in which the output takes only discrete values.

We now apply our Bayesian sparse reconstruction framework to neural networks, and show that this approach for principled adaptive Bayesian selection of network architecture without Gaussian approximations works well for ``shallow'' neural networks with a small number of hidden layers.
We use tanh activation functions for the nodes in the $L$ hidden layers, and a sigmoid activation function for the output
\begin{equation}
    \phi^{[L+1]}(x) = \mathrm{sigmoid}(x) = \frac{1}{1+\e^{-x}} = \frac{\e^x}{1+\e^x}\label{equ:sigmoid}.
\end{equation}
This conveniently maps the output $y$ into $[0,1]$, which is the range of the target signal in the numerical examples.

We use Gaussian priors on the neural network's weight parameters, as summarised in \Cref{tab:priors_nn}.
Due to the difficulty in selecting the priors' scale {\em a priori}, we use a hyperparameter $\sigma_w$ for the width of the Gaussian priors on the weights --- this can be marginalised out when calculating posterior inferences.
Following \citet{MacKay1995} we use a uniform prior on $\sigma_w^{-2}$, meaning
\begin{equation}
    \pi(\sigma_w) = \frac{3 \sigma_w^{-3}}{\sigma_{w,\min}^{-2} - \sigma_{w,\max}^{-2}}
    \label{equ:sigma_w_prior}
\end{equation}
where $\sigma_w > 0$.

\begin{table}
    \centering
    \caption{Priors on neural network parameters.
        Sorted priors have ordering enforced; see \citet[][Appendix A2]{Handley2015b} for more details.}%
    \label{tab:priors_nn}
    \begin{tabular}{lll}
    \toprule
    Parameter                       & Prior Type                                           & Prior Parameters                \\
    \midrule                            
    $L$                             & Uniform (integer)                                    & $\in \mathbb{Z} \cap [1, 2]      $ \\
    $N$                             & Uniform (integer)                                    & $\in \mathbb{Z} \cap [1, 10]     $ \\
    $\sigma_w$                      & Uniform in $\sigma_w^{-2}$~\eqref{equ:sigma_w_prior} & $\in [0.1, 10]                   $ \\
    output weights $w^{[L+1]}_{ij}$ & Sorted Gaussian\footnotemark{}                       & $\mu=0,\sigma=\sigma_w           $ \\
    other weights \& biases         & Gaussian                                             & $\mu=0,\sigma=\sigma_w           $ \\
    \bottomrule
    \end{tabular}
\end{table}\footnotetext{For neural networks with only one hidden layer, following \citet{MacKay1995}, priors on the weights leading to the output are further restricted to only be non-zero in the positive half of the Gaussian.
This exploits a symmetry in the parameter space as $\tanh(x)$ is symmetric under changes of sign in $x$.}

\subsection{Fitting 2-dimensional images with neural networks}

\begin{figure*}
	\centering
    \includegraphics[trim={0.35in 0 0 0}, clip]{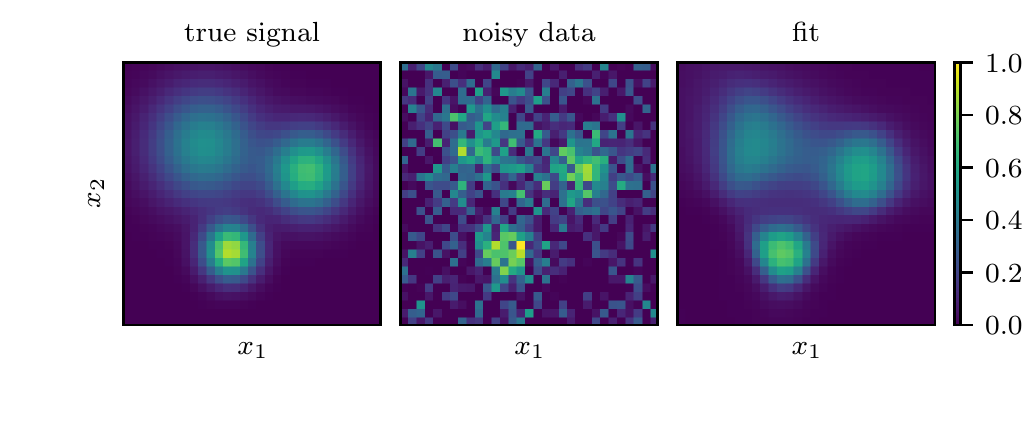}
    \includegraphics{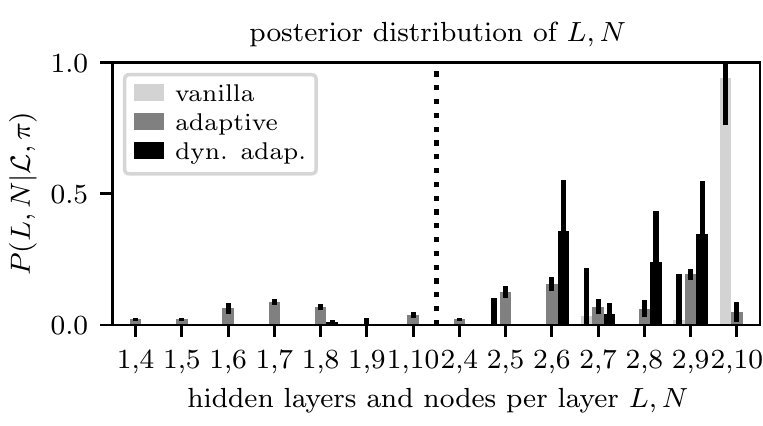}
    \caption{Fitting neural networks with the number of hidden layers $L$ and nodes per hidden layer $N$ determined through Bayesian inference.
    The first two colour plots show the true signal and the data, which includes added normally distributed $y$-errors with $\sigma_y=0.2$; these are the same as in \Cref{sub:multi_gg_2d_gg_2d_3}.
    The third colour plot shows the mean value of $y(\bm{x})$ from the posterior samples produced using the adaptive method with dynamic nested sampling.
    The bar plot displays the posterior distribution on $L,N$; values calculated using the vanilla method and the adaptive method without nested sampling are also included for comparison.
    Bars showing posterior probabilities for $N=1$, $N=2$ and $N=3$ are omitted for brevity as they contain negligible posterior mass for both $L=1$ and $L=2$.
    Adaptive results use a combined inference from 5 runs, each of which computes a full posterior on $L,N$ and uses 2,000 live points; adaptive runs using dynamic nested sampling have \dyPolyChord{} settings $n_\mathrm{init}=1,000$ and $\texttt{dynamic\_goal}=1$.
    Results for the vanilla method use 5 separate runs, each with 400 live points, to compute the evidence for each combination $L,N$.
    All runs use the setting $\texttt{num\_repeats}=250$}%
    \label{fig:multi_nn_adl_gg_2d}
    \includegraphics{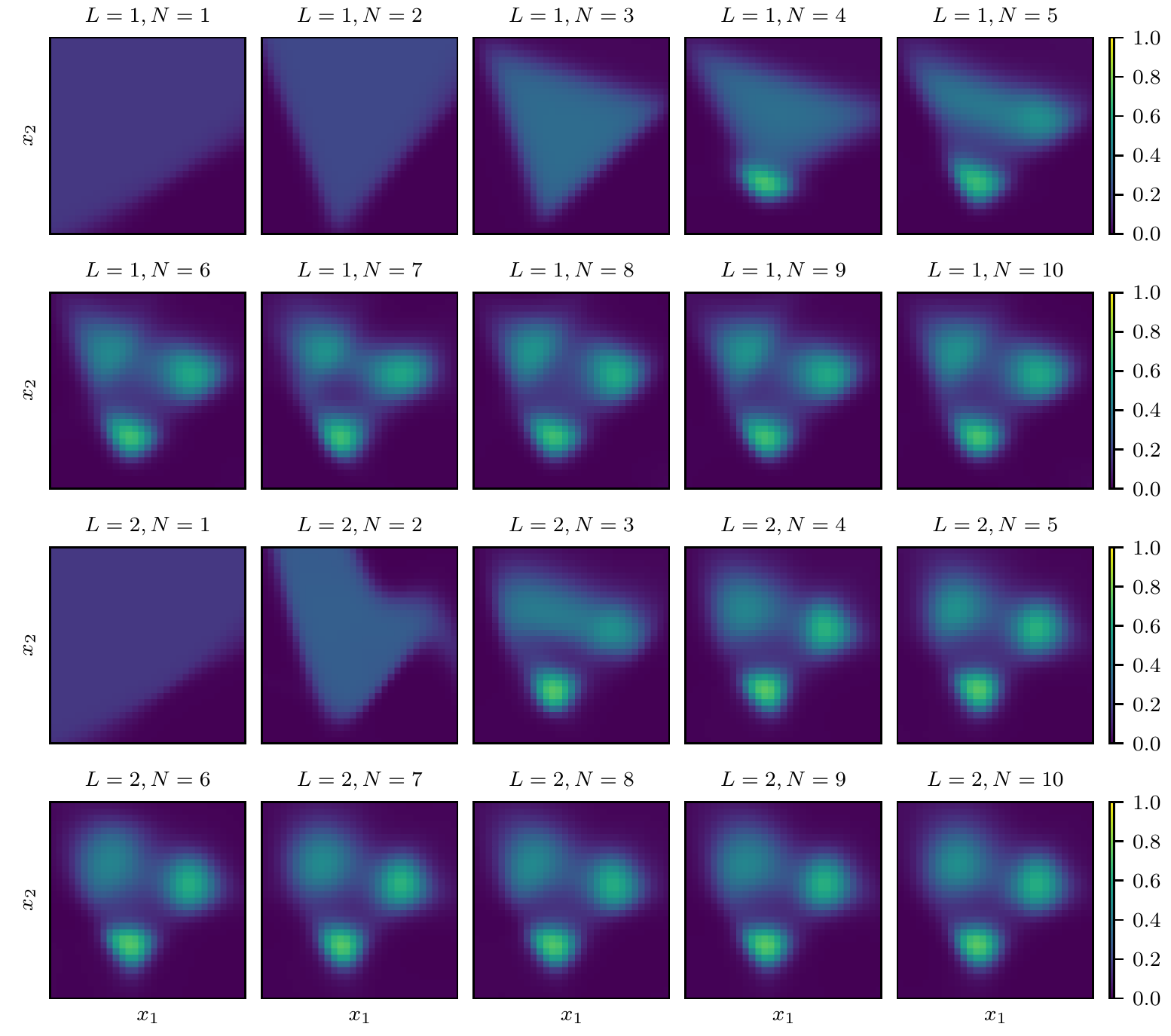}
    \caption{Fits from \Cref{fig:multi_nn_adl_gg_2d} conditioned on different numbers of hidden layers $L$ and nodes per hidden layer $N$. The plots use results from the vanilla method.}%
    \label{fig:split_nn_adl_gg_2d}
\end{figure*}

\Cref{fig:multi_nn_adl_gg_2d} shows signal reconstruction with neural networks, including Bayesian inference on the number of hidden layers $L$ and nodes per hidden layer $N$, using our Bayesian sparse reconstruction framework.
Readers who are less familiar with neural networks might expect them to struggle to fit the challenging data set (the same one used in \Cref{sub:multi_gg_2d_gg_2d_3}) using their tanh activation functions.
However we see that our approach yields good results, and the network is able to reconstruct the generalised Gaussians in the signal by overlaying 2-dimensional tanh functions from different nodes.
How it does this is illustrated in \Cref{fig:split_nn_adl_gg_2d}, which shows fits conditioned on different values of $L$ and $N$.
The first two rows with $L=1$ represent a network with a single hidden layer, and show how increasing $N$ allows the tanh functions to first create a triangle around the three maxima and then to represent the maxima themselves.
The second two rows use $L=2$; a comparison with the $L=1$ plots shows how the two hidden layer architecture allows more complex signal structure to be represented using a given value of $N$.
The posterior distribution of $L$ heavily favours two hidden layers, with the adaptive method using dynamic nested sampling giving $P(L=2|\mathcal{L},\pi) = 0.984 \pm 0.007$.

The network with $L=2$ hidden layers and $N=10$ nodes per hidden layer has 151 weight parameters plus the hyperparameter $\sigma_w$ and the integer parameters $L$ and $N$; the resulting parameter space is 154-dimensional, as well as highly multimodal and degenerate.
The default \PolyChord{} and \dyPolyChord{} settings for this dimensionality are $25 \times d = 3,850$ live points and $5 \times d = 770$, so it is not surprising that with the settings used our results show large inconsistencies in the calculated posterior distribution of $L$ and $N$ due to imperfect exploration of the parameter space \citep[see][for a detailed discussion]{Higson2018a}.
However our approach is still able to allocate almost all the posterior mass to $L,N$ combinations which are good fits for the data, leading to good results and demonstrating the robustness of the method.

Furthermore, neural network and basis function fits can be compared using the adaptive method.
For example one could include an additional integer parameter $T$, with values $T=1$ and $T=2$ representing fitting with 2-dimensional generalised Gaussians and with neural networks respectively.

\subsection{Application to astronomical images}\label{sec:nn_get_image}

We now apply neural networks to the Hubble Space Telescope eXtreme Deep Field images used in \Cref{sec:gg_2d_get_image}.
We find the adaptive selection of $L$ for these data sets strongly favours $L=2$ over $L=1$, so for brevity we show only results using 2 hidden layers.

\begin{figure*}
    \vspace{1cm}  
	\centering
    \begin{subfigure}{\linewidth}
        \includegraphics{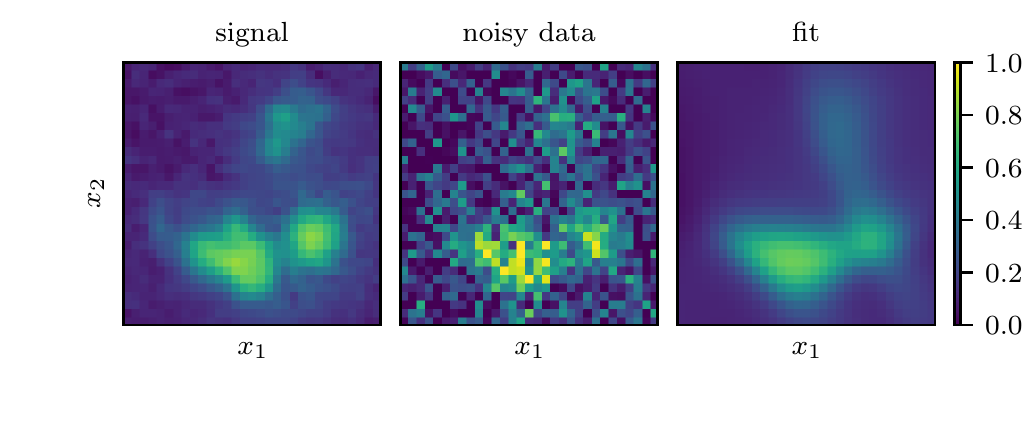}
        \includegraphics{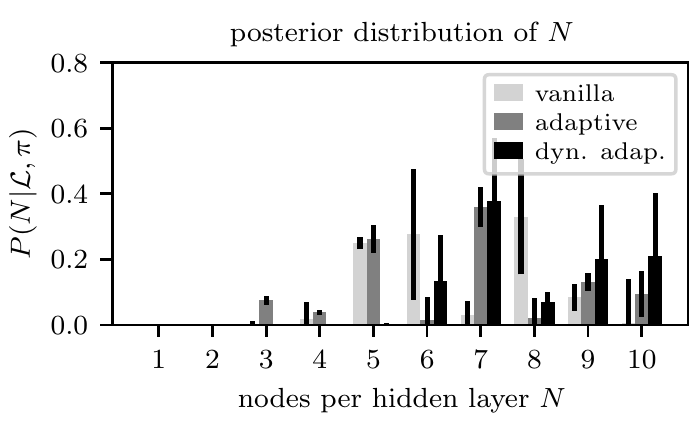}
        \subcaption{Image containing several galaxies.}%
        \label{sub:multi_nn_2l_get_image_2}
    \end{subfigure}
    \begin{subfigure}{\linewidth}
        \includegraphics{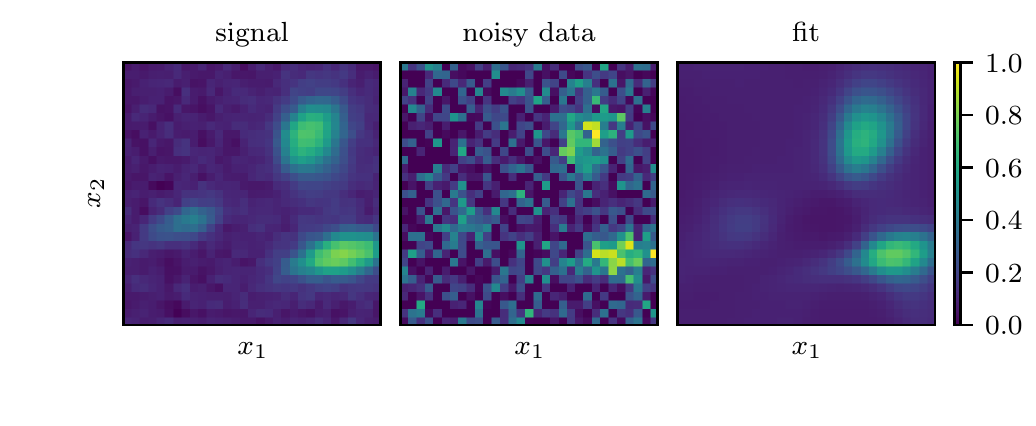}
        \includegraphics{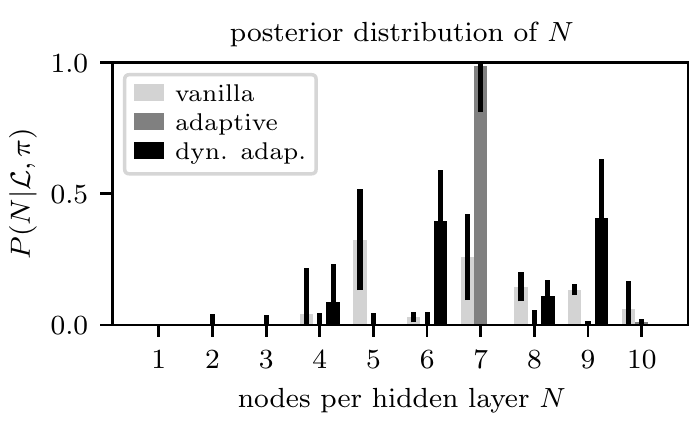}
        \caption{Another image containing several galaxies.}%
        \label{sub:multi_nn_2l_get_image_3}
    \end{subfigure}
    \caption{Fitting $32\times32$ images from the Hubble Space Telescope eXtreme Deep Field \citep{Illingworth2013} using neural networks with two hidden layers.
    In each row the 2 plots on the left show the true signal and the data, which includes added normally distributed $y$-errors with $\sigma_y=0.2$.
    The third column shows the mean value of $y$ from the posterior samples produced using the adaptive method with dynamic nested sampling.
    The bar plots display the posterior distribution for different numbers of nodes per hidden layer $N$; values calculated using the vanilla method and the adaptive method without dynamic nested sampling are also included for comparison.
    Adaptive results show a combined inference form 5 runs, each of which computes a full posterior on $N$ and uses 2,000 live points; adaptive runs using dynamic nested sampling have \dyPolyChord{} settings $n_\mathrm{init}=1,000$ and $\texttt{dynamic\_goal}=1$.
    Results for the vanilla method use separate runs, each with 400 live points, to compute the evidence for each value of $N$.
    All runs use the setting $\texttt{num\_repeats}=250$.}%
    \label{fig:multi_nn_2l_get_image}
    \includegraphics{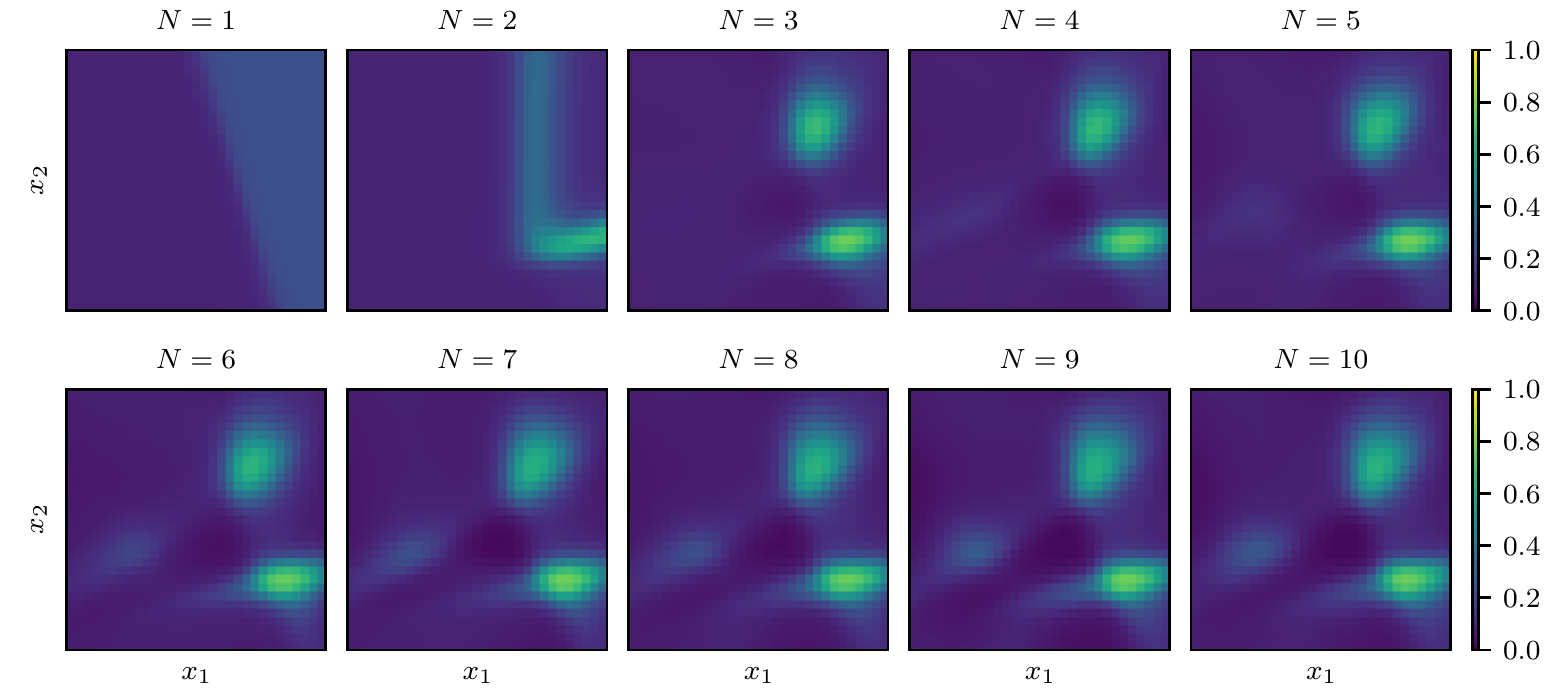}
    \caption{Fits of the data sets shown in \Cref{sub:multi_nn_2l_get_image_3} conditioned on different numbers of nodes per hidden layer $N$. The plots use results from the vanilla method.}%
    \label{fig:split_nn_2l_get_image}
    \vspace{1cm}  
\end{figure*}

\Cref{fig:multi_nn_2l_get_image} shows results from fitting neural networks with 2 hidden layers to the data used in \Cref{fig:multi_gg_2d_get_image}, with fits conditioned on specific values of $N$ shown in \Cref{fig:split_nn_2l_get_image}.
As for the 2-dimensional Gaussian basis functions, a visual assessment shows the neural networks are able to faithfully reconstruct the true image from the noisy data with good accuracy.
However the neural networks (with tanh activation functions) do not provide as natural a representation of the blob-shaped sources as the 2-dimensional generalised Gaussians, so the fits are not as good as those shown in \Cref{fig:multi_gg_2d_get_image,fig:split_gg_2d_get_image}.
Nevertheless the example provides a proof of principle, and the versatility of neural networks means they can be applied to a wide range of data sets using this technique.

The posterior distributions of the number of nodes in each hidden layer $N$, shown on the right of each row of plots in \Cref{fig:multi_nn_2l_get_image}, illustrate the number of nodes (degree of complexity of the model) which is justified by the astronomical image.
However, unlike the basis functions, the contributions of each individual node to the output fit is not readily interpretable.
As in the previous section, we find that the network's fits of the images are good --- despite inconsistencies in the posterior probabilities of different values of $N$ between the different methods due to the sampler imperfectly exploring the parameter space.
The posterior distribution of $N$ can be calculated more precisely using more computational resources (for example by increasing \PolyChord{} and \dyPolyChord's \numrepeats{} settings).

\section{Conclusion}

We have introduced Bayesian sparse reconstruction; a principled framework for signal reconstruction which allows the model's complexity to be determined by the data.
The Bayesian calculations naturally penalise over-complex models, and the priors can be used to specify the degree to which sparse solutions should be favoured.
Our approach performs well at fitting noisy 1- and 2-dimensional test data from mixture models, as well as reconstructing astronomical images.
We further showed the framework naturally applies to neural networks, where it allows Bayesian inference to be performed over the space of possible network architectures by treating the number of nodes and hidden layers as parameters.

While the techniques described in this paper are computationally expensive (see \Cref{app:core_hours} for details of the compute used to produce our results), we show that they are now feasible in the low data regime with current software and are capable of producing excellent results.
Furthermore, we intend this paper to provide a proof of principle for future application of our approach to larger data sets, when advances in numerical techniques and increases in computational power make this feasible.

\section*{Acknowledgements}

This work was performed using the Darwin Supercomputer of the University of Cambridge High Performance Computing Service (\href{http://www.hpc.cam.ac.uk/}{\url{http://www.hpc.cam.ac.uk/}}), provided by Dell Inc.\ using Strategic Research Infrastructure Funding from the Higher Education Funding Council for England and funding from the Science and Technology Facilities Council.





\bibliographystyle{mnras}
\bibliography{library} 



\appendix

\section{Code}\label{app:code}

The code used to make the results and plots in this paper can be downloaded at \href{https://github.com/ejhigson/bsr}{\url{https://github.com/ejhigson/bsr}}.

\section{Computational resources used}\label{app:core_hours}

\Cref{tab:core_hours} shows the approximate number of core hours used for each calculation in this paper, and is intended to provide a rough guide to the computational cost of our method.
We used the CDS3 Peta4 cluster, which has 2.6GHz 16-core Intel Xeon Skylake 6142 processors (2 processors and 32 cores per node).
Note that the number of core hours used can vary significantly when the same calculation is repeated.

Vanilla and adaptive calculations use \PolyChord{} and dynamic adaptive calculations use \dyPolyChord{}.
\dyPolyChord{} performs dynamic nested sampling by saving and resuming \PolyChord{} runs; this is not yet parallelised in the current version of \PolyChord{} and can become a bottleneck when running with large numbers of processes, increasing the amount of core hours required for this method.
We intend this process to be more computationally efficient in future dynamic nested sampling software.
All calculations use C\texttt{++} likelihoods except the adaptive and dynamic adaptive selection of $T$ in \Cref{fig:multi_ta_1d_ta_1d}, which were run using a Python likelihood and consequently required more computation time.
The code used can be downloaded from the link in Appendix~\ref{app:code}.

When fitting the same basis functions to different data sets, reconstructing more complex signals requires more computation --- this can be seen in \Cref{tab:core_hours} for \Cref{sub:multi_gg_1d_gg_1d_1,sub:multi_gg_1d_gg_1d_2,sub:multi_gg_1d_gg_1d_3} and \Cref{sub:multi_ta_1d_ta_1d_1,sub:multi_ta_1d_ta_1d_2,sub:multi_ta_1d_ta_1d_3}.

\begin{table}
\centering
\caption{Approximate numbers of core hours used per calculation for results shown in the paper; these were run on the CDS3 Peta4 cluster, which has 2.6GHz 16-core Intel Xeon Skylake 6142 processors (2 processors and 32 cores per node).
         For adaptive results, each calculation is a single nested sampling run.
         For vanilla runs a calculation involves a separate nested sampling run for each value of $N$ --- the values of the table show the total core hours used by these.
         For calculations fitting basis functions to 1-dimensional signals (\Cref{fig:multi_gg_1d_gg_1d,fig:multi_ta_1d_ta_1d,fig:multi_adfam_gg_ta_1d}) \numrepeats=100, vanilla runs use 200 live points and adaptive runs use 1,000.
         For calculations fitting 2-dimensional images (\Cref{fig:multi_gg_2d_gg_2d,fig:multi_gg_2d_get_image,fig:multi_nn_adl_gg_2d,fig:multi_nn_2l_get_image}) \numrepeats=250, vanilla runs use 400 live points and adaptive runs use 2,000.
         Note that plots of results all use combined inferences from 5 calculations.}\label{tab:core_hours}
\begin{tabular}{llll}
\toprule
                                         &    vanilla &   adaptive & dynamic adaptive \\
\midrule
\multicolumn{4}{l}{Fitting 1$d$ generalised Gaussians (\Cref{fig:multi_gg_1d_gg_1d})} \\
\Cref{sub:multi_gg_1d_gg_1d_1}           &          1 &          1 &                3 \\
\Cref{sub:multi_gg_1d_gg_1d_2}           &        1.5 &        1.5 &                4 \\
\Cref{sub:multi_gg_1d_gg_1d_3}           &          2 &          2 &                4 \\
\midrule
\multicolumn{4}{l}{Fitting 1$d$ tanhs (\Cref{fig:multi_ta_1d_ta_1d})}                 \\
\Cref{sub:multi_ta_1d_ta_1d_1}           &          1 &          1 &                3 \\
\Cref{sub:multi_ta_1d_ta_1d_2}           &        1.5 &        1.5 &                4 \\
\Cref{sub:multi_ta_1d_ta_1d_3}           &          3 &          3 &                5 \\
\midrule
\multicolumn{4}{l}{Fitting 1$d$ basis functions with adaptive $T$ (\Cref{fig:multi_adfam_gg_ta_1d})}     \\
\Cref{sub:multi_adfam_gg_ta_1d_ta_1d_1}  &          2 &         20 &                20 \\
\Cref{sub:multi_adfam_gg_ta_1d_gg_1d_1}  &          2 &         20 &                20 \\
\midrule
\multicolumn{4}{l}{Fitting 2$d$ generalised Gaussians (\Cref{fig:multi_gg_2d_gg_2d,fig:multi_gg_2d_get_image})}     \\
\Cref{sub:multi_gg_2d_gg_2d_1}           &          5 &         7 &                50 \\
\Cref{sub:multi_gg_2d_gg_2d_2}           &         10 &        14 &                70 \\
\Cref{sub:multi_gg_2d_gg_2d_3}           &         12 &        20 &                80 \\
\Cref{sub:multi_gg_2d_get_image_1}       &          8 &        26 &                70 \\
\Cref{sub:multi_gg_2d_get_image_2}       &         11 &        60 &               100 \\
\Cref{sub:multi_gg_2d_get_image_3}       &         10 &        40 &                80 \\
\midrule
\multicolumn{4}{l}{Fitting neural networks with adaptive $L$ (\Cref{fig:multi_nn_adl_gg_2d})}     \\
\Cref{fig:multi_nn_adl_gg_2d}            &        150 &        200 &              300 \\
\midrule
\multicolumn{4}{l}{Fitting neural networks with 2 hidden layers (\Cref{fig:multi_nn_2l_get_image})}     \\
\Cref{sub:multi_nn_2l_get_image_2}       &        100 &        100 &              200 \\
\Cref{sub:multi_nn_2l_get_image_3}       &        100 &        100 &              200 \\
\bottomrule
\end{tabular}
\end{table}

\section{Additional numerical results}\label{app:anr}

This Appendix contains details of the parameters of the mixture models used to generate true signals in the numerical examples, as well as tables comparing the computational efficiency of results calculated through the adaptive and vanilla methods.

\subsection{Parameters for test signals}

\Cref{tab:gg_1d_data_args,tab:ta_1d_data_args,tab:gg_2d_data_args} show the parameters of the mixture models used for the signals in Figures~\ref{fig:multi_gg_1d_gg_1d},~\ref{fig:multi_ta_1d_ta_1d} and~\ref{fig:multi_gg_2d_gg_2d} respectively.

\begin{table}
    \centering
    \caption{Parameters for the sum of 1-dimensional generalised Gaussian basis functions~\eqref{equ:gg_1d} from which the data shown in \Cref{fig:multi_gg_1d_gg_1d} was sampled.}%
    \label{tab:gg_1d_data_args}
    \begin{tabular}{lllll}
    \toprule
    \# functions  & $a$  & $\mu$ & $\sigma$  & $\beta$ \\
    \midrule
    1             & 0.75 & 0.4   & 0.3       & 2       \\
    \midrule
    2             & 0.2  & 0.4   & 0.6       & 5       \\
                  & 0.55 & 0.4   & 0.2       & 4       \\
    \midrule
    3             & 0.2  & 0.4   & 0.6       & 5       \\
                  & 0.35 & 0.6   & 0.07      & 2       \\
                  & 0.55 & 0.32  & 0.14      & 6       \\
    \bottomrule
    \end{tabular}
\end{table}

\begin{table}
    \centering
    \caption{Parameters for the sum of 1-dimensional tanh basis functions~\eqref{equ:ta_1d} from which the data shown in \Cref{fig:multi_ta_1d_ta_1d} was sampled.}%
    \label{tab:ta_1d_data_args}
    \begin{tabular}{llll}
    \toprule
    \# functions  & $a$  & $b$   & $w$   \\
    \midrule
    1             & 0.8  & 0     & 1.5     \\
    \midrule
    2             & 0.7  & -1    & 3     \\
                  & 0.9  & 2     & -3     \\
    \midrule
    3             & 0.6  & -7    & 8     \\
                  & 1    & -1    & 3    \\
                  & 1.4  &  2    & -3    \\
    \bottomrule
    \end{tabular}
\end{table}

\begin{table}
    \centering
    \caption{Parameters for the sum of two-dimensional generalised Gaussian basis functions~\eqref{equ:gg_2d} from which the data shown in \Cref{fig:multi_gg_2d_gg_2d} was sampled.}%
    \label{tab:gg_2d_data_args}
    \begin{tabular}{lllllllll}
    \toprule
    \# functions  & $a$  & $\mu_1$ & $\mu_2$  & $\sigma_1$ & $\sigma_2$ & $\beta_1$ & $\beta_2$ & $\Omega$ \\
    \midrule
    1             & 0.8  & 0.6     & 0.6      & 0.1        & 0.2        & 2         & 2         & $\pi/10$ \\
    \midrule
    2             & 0.5  & 0.5     & 0.4      & 0.4        & 0.2        & 2         & 2         & 0        \\
                  & 0.8  & 0.5     & 0.6      & 0.1        & 0.1        & 2         & 2         & 0        \\
    \midrule
    3             & 0.5  & 0.3     & 0.7      & 0.2        & 0.2        & 2         & 2         & 0        \\
                  & 0.7  & 0.7     & 0.6      & 0.15       & 0.15       & 2         & 2         & 0        \\
                  & 0.9  & 0.4     & 0.3      & 0.1        & 0.1        & 2         & 2         & 0        \\
    \bottomrule
    \end{tabular}
\end{table}

\subsection{Efficiency gain results}

\Cref{tab:gg_1d_gg_1d_results,tab:gg_2d_gg_2d_results} show numerical values for the mean fit at the centre of the signal's domain for \Cref{fig:multi_gg_1d_gg_1d,fig:multi_gg_2d_gg_2d}, as well as estimates of the efficiency gain~\eqref{equ:efficiency_gain_bsr} from the adaptive method (with and without dynamic nested sampling) compared to the vanilla method.
Efficiency gains reported use results' estimated variation, calculated from bootstrap resampling using the \nestcheck{} package \citep{Higson2018nestcheck}.

Bootstrap resampling allows the variation of results due to the stochasticity of the nested sampling algorithm to be determined accurately without the need to repeat the calculation many times.
However it assumes that the nested sampling algorithm was performed perfectly; for some special cases this is possible \citep[see for example][]{Higson2018perfectns}, but in practice for challenging posterior distributions there may be additional errors --- for example due to software producing correlated samples, or missing a mode in a multimodal posterior.
These additional errors are discussed in detail in \citet{Higson2018a}, and can be reduced by changing the software settings; for \PolyChord{} and \dyPolyChord{} this entails increasing the \numrepeats{} setting and/or the number of live points.
Diagnostics provided by \nestcheck{} indicate the presence of such additional variation in our results; it also explains how estimates of the fit $y(0.5;\bm{\theta})$ and $y(0.5,05;\bm{\theta})$ using different methods (shown in \Cref{tab:gg_1d_gg_1d_results,tab:gg_2d_gg_2d_results} respectively) sometimes differ by slightly more than would be expected from their bootstrap uncertainties.
As the posterior is more challenging and complex in the adaptive method than the vanilla method, this is likely to mean the efficiency gain observed in practice is lower than the estimates using the bootstrap estimates of variation with the settings we use. However we include it as a rough estimate and an indication of the efficiency gain which could be achieved with more live points and/or a higher \numrepeats{} setting.

\begin{table}
\centering
\caption{Numerical values for the accuracy of the mean fit at $x=0.5$ using the data and nested sampling runs shown in \Cref{fig:multi_gg_1d_gg_1d}.
The columns show results for the vanilla and adaptive methods using standard nested sampling and the adaptive method using dynamic nested sampling.
For each data set, the first two rows show the total number of samples used by the nested sampling runs, and the mean value of $y(0.5;\bm{\theta})$.
The next two rows show the efficiency gain~\eqref{equ:efficiency_gain_bsr} of the adaptive method with and without dynamic nested sampling; these are calculated using estimates of the standard deviation of results from bootstrap resampling 100 bootstrap replications.
The numbers in brackets show $1\sigma$ errors on the final digit.}\label{tab:gg_1d_gg_1d_results}
\begin{tabular}{lllll}
\toprule
           &    vanilla &    adaptive & dynamic adaptive \\
\midrule
\multicolumn{4}{l}{Data from 1 generalised Gaussian (shown in \Cref{sub:multi_gg_1d_gg_1d_1})}     \\
\# samples           &    128,719 &     114,507 &          116,867 \\
$y(0.5;\bm{\theta})$ &  0.6526(4) &   0.6517(3) &        0.6527(1) \\
efficiency gain      &         &      1.7(4) &            46(9) \\
\midrule
\multicolumn{4}{l}{Data from 2 generalised Gaussians (shown in \Cref{sub:multi_gg_1d_gg_1d_2})}     \\
\# samples &    128,052 &     133,061 &          131,637 \\
$y(0.5;\bm{\theta})$ &  0.6340(3) &  0.6351(1) &        0.6338(1) \\
efficiency gain &         &       14(3) &           4.3(9) \\
\midrule
\multicolumn{4}{l}{Data from 3 generalised Gaussians (shown in \Cref{sub:multi_gg_1d_gg_1d_3})}     \\
\# samples &    152,516 &     171,853 &          173,959 \\
$y(0.5;\bm{\theta})$ &  0.4040(3) &   0.4029(2) &        0.4072(2) \\
efficiency gain &         &      2.5(5) &           2.7(6) \\
\bottomrule
\end{tabular}
\end{table}

\begin{table}
\centering
\caption{Numerical values for the accuracy of the mean fit at $\bm{x}=(0.5,0.5)$ using the data and nested sampling runs shown in \Cref{fig:multi_gg_2d_gg_2d}.
The columns show results for the vanilla and adaptive methods using standard nested sampling and the adaptive method using dynamic nested sampling.
For each data set, the first two rows show the total number of samples used by the nested sampling runs, and the mean value of $y(0.5,0.5;\bm{\theta})$ produced.
The next two rows show the efficiency gain~\eqref{equ:efficiency_gain_bsr} of the adaptive method with and without dynamic nested sampling; these are calculated using estimates of the standard deviation of results from bootstrap resampling.
The numbers in brackets show $1\sigma$ errors on the final digit.}\label{tab:gg_2d_gg_2d_results}
\begin{tabular}{llll}
\toprule
     &    vanilla &   adaptive & dynamic adaptive \\
\midrule
\multicolumn{4}{l}{Data from 1 2$d$ generalised Gaussian (shown in \Cref{sub:multi_gg_2d_gg_2d_1})}     \\
\# samples                &    377,360 &    324,294 &          322,799 \\
$y(0.5,0.5;\bm{\theta})$  &  0.3353(3) &  0.3360(2) &       0.3359(1) \\
efficiency gain           &            &     3.7(7) &            16(3) \\
\midrule
\multicolumn{4}{l}{Data from 2 2$d$ generalised Gaussians (shown in \Cref{sub:multi_gg_2d_gg_2d_2})}     \\
\# samples                &    476,932 &    520,691 &          503,380 \\
$y(0.5,0.5;\bm{\theta})$  &  0.6850(5) &  0.6853(2) &        0.6856(1) \\
efficiency gain           &            &      10(2) &            12(2) \\
\midrule
\multicolumn{4}{l}{Data from 3 2$d$ generalised Gaussians (shown in \Cref{sub:multi_gg_2d_gg_2d_3})}     \\
\# samples               &    562,830 &    652,359 &          656,852 \\
$y(0.5,0.5;\bm{\theta})$ &  0.1435(1) &  0.1438(1) &       0.1437(1) \\
efficiency gain          &            &     1.2(2) &           4.3(9) \\
\bottomrule
\end{tabular}
\end{table}


\bsp\label{lastpage}
\end{document}